\documentclass[journal]{IEEEtran}
\usepackage{cite}

\ifCLASSINFOpdf
  \usepackage[pdftex]{graphicx}
\else
\fi
\usepackage[caption=false, font=footnotesize]{subfig}

\usepackage{amsmath,mleftright}
\usepackage{amssymb}
\DeclareMathOperator*{\argmin}{argmin}

\usepackage{xparse}
\NewDocumentCommand{\evalat}{sO{\big}mm}{%
  \IfBooleanTF{#1}
   {\mleft. #3 \mright|_{#4}}
   {#3#2|_{#4}}%
}

\usepackage{lipsum}
\usepackage{array,multirow}
\usepackage{url}

\begin{document}
%
\title{Calibration-Free Positioning Technique Using Wi-Fi Ranging and Built-in Sensors of Mobile Devices}
%
%
%

\author{Jeongsik~Choi,~\IEEEmembership{Member,~IEEE} and Yang-Seok~Choi
\thanks{Copyright (c) 2020 IEEE. Personal use of this material is permitted. However, permission to use this material for any other purposes must be obtained from the IEEE by sending a request to pubs-permissions@ieee.org.}
\thanks{J.~Choi is with the Intel Labs, Intel Corporation, Santa Clara, CA 95054, USA (e-mail: jeongsik.choi@intel.com).}
\thanks{Y.-S.~Choi is with the Intel Labs, Intel Corporation, Hillsboro, OR 97124, USA (e-mail: yang-seok.choi@intel.com).}
}

\IEEEaftertitletext{\vspace{-1.1\baselineskip}}

%
%

\markboth{Accepted for publication in IEEE Internet of Things Journal}
{}
%



\maketitle

\begin{abstract}
As positioning solutions integrate multiple components to improve accuracy, the number of parameters that require calibration has increased.
This paper studies a calibration-free positioning technique using Wi-Fi ranging and pedestrian dead reckoning (PDR), where every parameter in the system is optimized in real-time.
This significantly decreases the time and effort required to perform manual calibration procedures and enables the positioning solution to achieve robust performance in various situations.
Additionally, this paper studies an efficient way of performing irregular Wi-Fi ranging procedures to improve battery life and network performance of mobile devices.
The positioning performance of the proposed method was verified using a real-time Android application on several mobile devices under a large indoor office environment.
Without any calibration, the proposed method achieved up to 1.38~m average positioning accuracy for received signal strength (RSS)-based ranging scenarios, which differs only by 30~cm from the benchmark assuming perfect calibration.
In addition, the proposed method achieved up to 1.04~m accuracy for round trip time (RTT)-based ranging scenarios with a 40~MHz bandwidth configuration, which differs only by 10~cm from the benchmark.
\end{abstract}


%
\IEEEpeerreviewmaketitle

\section{Introduction}

\IEEEPARstart{L}{ocating} mobile devices is one of essential technologies for various location-based services.
To provide seamless and precise positioning results, numerous attempts have been introduced in literature~\cite{4796924, 7039067, correa_sen_17, 8409950, 8852722}.
In particular, many approaches have been introduced positioning solutions that rely only on the built-in components of devices, such as Wi-Fi, Bluetooth, GPS, accelerometer, gyroscope, magnetometer, and so on~\cite{8007254,Yu_2019,8839041,8911751,8924707,Tian2014,Diaz2015,6987239,6971168,Wang2018,8455482,8385119,Zhang2018,8756098,Xu2019,Liu2012,6714451,6817916,7935650,8756276,8854290,9043567}.
The wireless components are typically used to estimate the absolute position of the device by utilizing fixed reference nodes in the vicinity~\cite{8007254,Yu_2019,8839041,8911751,8924707,8854290, 9043567}, while sensors are used to obtain changes in the relative position of the device over time~\cite{Tian2014,Diaz2015,6987239,6971168,Wang2018,8455482,8385119,Zhang2018,8756098,Xu2019}.
In addition, many positioning solutions have integrated both wireless components and sensors to achieve more accurate positioning performance~\cite{Liu2012,6714451,6817916,7935650,8756276}.

Among wireless components, Wi-Fi has been widely used for indoor positioning purposes because many indoor sites are already equipped with a sufficient number of Wi-Fi access points (APs) that can be used as reference nodes.
The received signal strength (RSS) is one of the most popular sources for Wi-Fi-based positioning solutions.
With RSSs from multiple nearby APs, the distance to each AP can be secured, and subsequently, the location of the device is obtained by applying the trilateration techniques~\cite{Wang2003AnIW,5425237,5558044,5766644,7935650,8839041}.
Moreover, the RSS is also widely used in the Wi-Fi fingerprinting technique, which tabulates RSSs from nearby APs measured at various locations in the area of interests and uses this database to locate a device~\cite{832252,1047316,horus05,Liu2012,8007254}.

The IEEE 802.11-2016 standard defined a round trip time (RTT)-based ranging protocol called fine timing measurement (FTM) to enhance the positioning capability though accurate ranging~\cite{7786995}.
Because this protocol measures distance in the form of propagation time multiplied by the speed of light, the ranging accuracy primarily depends on how precisely the direct path is detected among multi-path components.
Hence, the FTM protocol provides accurate ranging results for a line-of-sight (LOS) propagation environment or when using a wide bandwidth.
However, a known challenge related to the FTM protocol is that raw distance measurements are distorted due to many factors such as cable length and timing offset of wireless packets~\cite{ibrahim_mobicom_18,8839041,8911824,8911751,8924707}.
Therefore, a calibration procedure is essential for the FTM protocol to operate accurately.

Unlike wireless components, built-in sensors are less affected by the external environment and have been widely used to estimate the movement trajectory of a device.
A simple method to estimate the trajectory is to perform a double integral of acceleration on each axis~\cite{8455482, Zhang2018}.
However, the low-cost sensors equipped on mobile devices are generally not accurate enough to estimate the exact trajectory, unless the estimated results are corrected periodically, for example through zero velocity update (ZUPT)~\cite{Khairi_14}.
Hence, many studies have focused on an alternative approach called pedestrian dead reckoning (PDR), which estimates the traveled distance by counting the number of steps of the user and combines it with a separately estimated heading angle to obtain the trajectory of the device~\cite{Liu2012,Tian2014,Diaz2015,6987239,6971168,7935650,Wang2018,8385119,8756098,8756276,Xu2019}.

In this paper, we focus on a positioning solution based on a Wi-Fi ranging and PDR technique.
As the two approaches are fused to produce more accurate positioning results, the number of configurable parameters increases.
In addition, the calibration process is very time consuming because the optimal set of parameters may vary from device to device or even from site to site.
To efficiently optimize all the parameters without human intervention, this paper proposes an online calibration technique that adaptively updates every parameter as the positioning application is operating.
The only requirement for this application is to determine the coordinates of the installed APs, which can be conducted using a one-time effort when they are installed or by using a semi-automated method introduced in~\cite{8911824}.

This paper addresses the following contributions:
\begin{enumerate}
    \item[1)] \emph{Online Parameter Calibration}: For RSS-based ranging scenarios, a specific signal attenuation pattern for each site (e.g., pathloss model) should be investigated.
    Similarly, an optimal calibration strategy should be known for RTT-based positioning scenarios to correct distortions in raw distance measurements.
    The PDR technique also has some trainable parameters such as user step length, initial position, and reference heading direction.
    The proposed method automatically optimizes all the parameters in real-time.
    
    \item[2)] \emph{Limited Wi-Fi Ranging Procedures}: The ranging procedure consumes a remarkable amount of power, as a Wi-Fi device actively scans multiple channels to collect RSSs or exchanges wireless packets multiple times to acquire RTT-related measurements. Furthermore, the network performance may be degraded when the Wi-Fi device is busy performing ranging procedures. Therefore, this paper studies an efficient method to balance positioning accuracy and battery efficiency by controlling the number of Wi-Fi ranging procedures.
    
    \item[3)] \emph{Heading Estimation without the Magnetometer}: According to the experimental results, estimated heading direction using the magnetometer widely fluctuates in an indoor environment because of the distortion of the magnetic field. For this reason, this study minimizes the use of the magnetometer and estimates the heading direction using Wi-Fi ranging and PDR.

    \item[4)] \emph{Extensive Experimentation in a Real Environment}: To verify the effectiveness of the proposed method, we implemented a real-time positioning Android application for both RSS and RTT-based ranging scenarios. Using this application, we conducted extensive experimental campaigns using eight different mobile devices in a large indoor office environment with 69 Wi-Fi APs. 
\end{enumerate}

The rest of the paper is organized as follows. A literature survey and system model are introduced in Sections II and III, respectively.
In section IV, the online calibration method is proposed, and the performance of this method is verified in Section V, and the paper is concluded.

Notation: $\mathbf A\in\mathbb{R}^{N\times M}$ represents an $N\times M$ real matrix (or vector) whose $(i, j)$-th element is denoted by $[\mathbf A]_{(i,j)}$. $\mathbf I_{N\times N} \in \mathbb{R}^{N\times N}$ and $\mathbf 0_{N \times M} \in \mathbb{R}^{N\times M}$ indicate the identity and the zero matrices, respectively.
$\mathbf A=\mbox{diag}(a_1, ..., a_N) \in \mathbb{R}^{N\times N}$ represents the diagonal matrix with diagonal elements $a_1, ..., a_N$. For a vector $\mathbf a \in \mathbb{R}^{N\times 1}$, $\lVert\mathbf a\rVert = \sqrt{\mathbf a^T \mathbf a}$ denotes the $l$2-norm of the vector. The inverse, transpose, and expectation operators are denoted by $(\cdot)^{-1}$, $(\cdot)^T$, and $E[\cdot]$, respectively.

\section{Related Works}

The pathloss model has been widely used for ranging using RSS measurements~\cite{Wang2003AnIW,5558044,5766644,7935650,8839041,8924707}.
This model adjusts the ranging strategy for each site and device using two trainable parameters, which are the RSS at a reference distance and the pathloss exponent (PLE).
In addition, a polynomial-based approach~\cite{5425237} and a neural network-based approach~\cite{8839041} are also introduced in literature to learn more flexible relationships between distance and RSS.
Irrespective of which ranging model is used, the parameters in the model should be selected carefully for each scenario. 
Such a calibration process is typically conducted by manually measuring the RSS at various distances from an AP (or multiple APs).

To minimize the time and effort for collecting training data to optimize the ranging model, online calibration techniques have been studied in~\cite{4027597, res_sen_11, 6509479}.
These techniques update the ranging parameters while the location of a device is estimated.
Another approach for minimizing human intervention in optimizing the ranging strategy is an unsupervised learning technique, which utilizes unlabeled data generated when users are using the positioning application~\cite{8839041}.
This technique focuses primarily on the site survey perspective, as it estimates optimal parameters that produce an excellent overall performance for a specific site, instead of updating parameters each time.
Furthermore, this technique also estimates the different characteristics across APs or devices.

The initial positioning studies using the FTM protocol were introduced in~\cite{ipin16, intel17, 8579543}.
In particular, fundamental operations of the ranging procedure and signal processing techniques such as packet exchange and clock drift correction were discussed in these papers.
As devices supporting the FTM protocol were introduced in the market, many studies verified the ranging and positioning performances of this new ranging method~\cite{ibrahim_mobicom_18,8756276,8839041,8911824,8911751,8924707}.
Almost every study addressed the issue related to the distortion of raw distance measurements using the FTM protocol, and positioning solutions in these studies performed a prior manual calibration procedure to remove such a distortion.

Two approaches have been used to estimate the trajectory of the device using sensors. The strapdown system conducts a double integral of acceleration on each axis to estimate the traveled distances along each axis~\cite{8455482, Zhang2018}.
However, as noise in sensor readings accumulates rapidly, the estimated distance easily diverges after a certain amount of time.
The second approach, the step and heading system, also known as the PDR technique, estimates the distance by multiplying the predefined step length and number of a user's detected steps~\cite{Liu2012,Tian2014,Diaz2015,6987239,6971168,7935650,Wang2018,8385119,8756098,8756276,Xu2019}.
Although the PDR technique can be free from the problem of noise accumulation, its positioning performance depends on many factors such as the accuracy of step detection, heading estimation, etc.
In addition, many approaches to estimate accurate step length for each user have been studied in literature~\cite{Zij_2004, 5646888, 6987239}.

\section{System Model}

\begin{figure}
    \centering
    \includegraphics[width=0.20\textwidth]{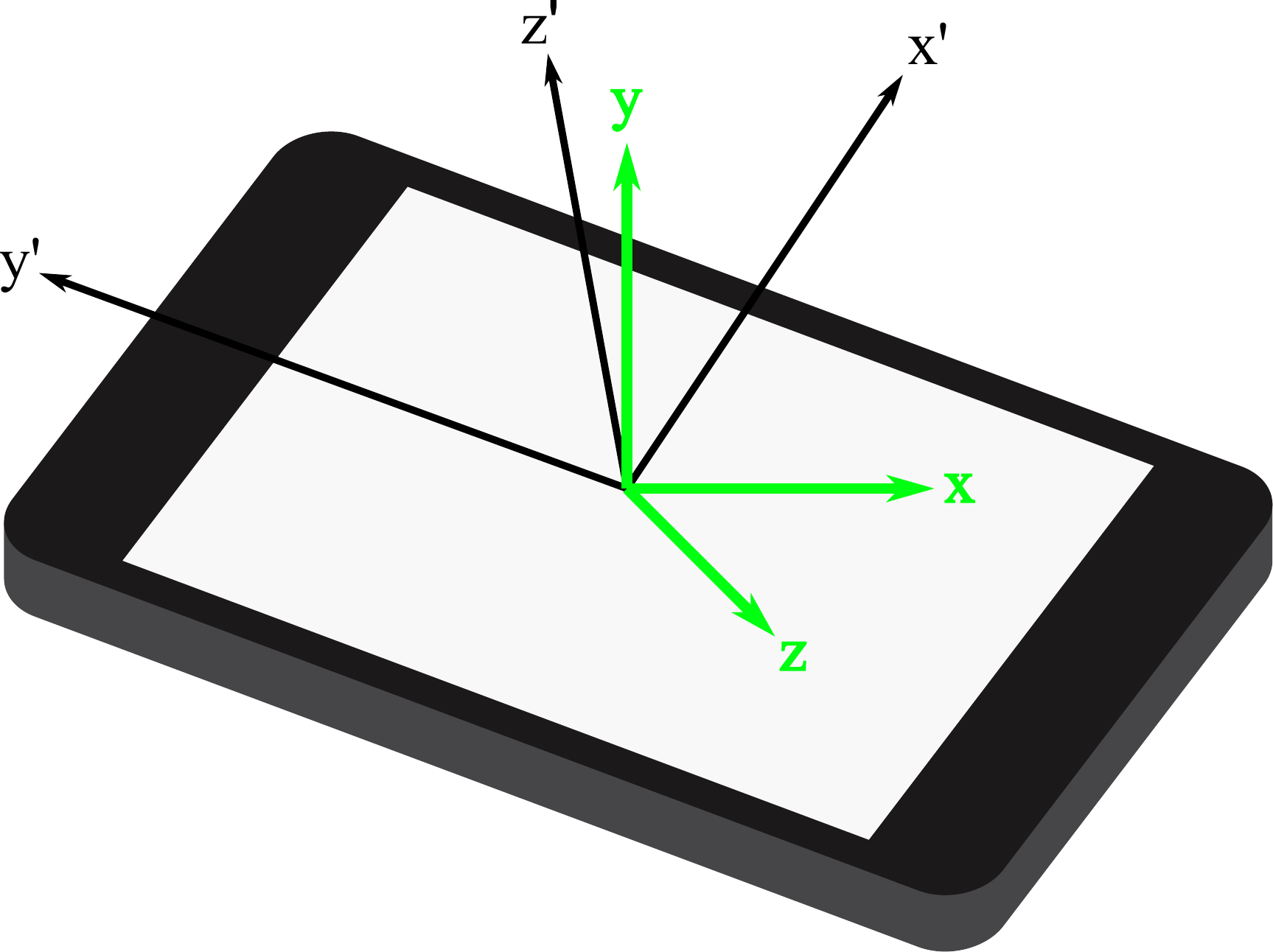}
    \caption{Two coordinate systems: the global coordinate system with x-, y-, and z-axes and the local coordinate system with x$'$-, y$'$-, and z$'$-axes.}
    \label{fig_coord}
\end{figure}

Fig.~\ref{fig_coord} illustrates two coordinate systems used in this paper. The global coordinate system (GCS) is a reference coordinate system for positioning, where the x-, y-, and z-axes point in the East, North, and up (ENU) directions, respectively, relative to the device's position on the Earth's surface.
The other system, the local coordinate system (LCS) is related to the device's orientation, where the x$'$- and y$'$-axes point in the right and upward directions of the device and the z$'$-axis points perpendicularly to the screen.
Every sensor reading is reported on the LCS, whereas the positioning results are yielded with respect to the GCS.

We consider a positioning scenario where every AP and the device are located at a similar altitude, for instance, on the same floor in a building.
Therefore, we primarily focus on an x-y plane of the GCS. We denote $\mathbf p = [x, y]^T$ and $\mathbf p_n = [x_n, y_n]^T$ as the position of the device and $n$-th AP on the positioning plane, respectively.
We assume that the positions of all APs are known as an essential requirement for range-based positioning techniques.
With this assumption, any difference between APs, such as transmission power, antenna gain, or mismatches in the FTM protocol, can be corrected as long as a user (e.g., a system operator) merely moves around the area~\cite{8839041}.
Thus, we can reasonably assume that all APs have the same characteristics.
In the remainder of this section, we briefly introduce the Wi-Fi ranging and PDR models.

\subsection{RSS and RTT-Based Wi-Fi Ranging Model}

\begin{figure}
    \centering
    \includegraphics[width=0.42\textwidth]{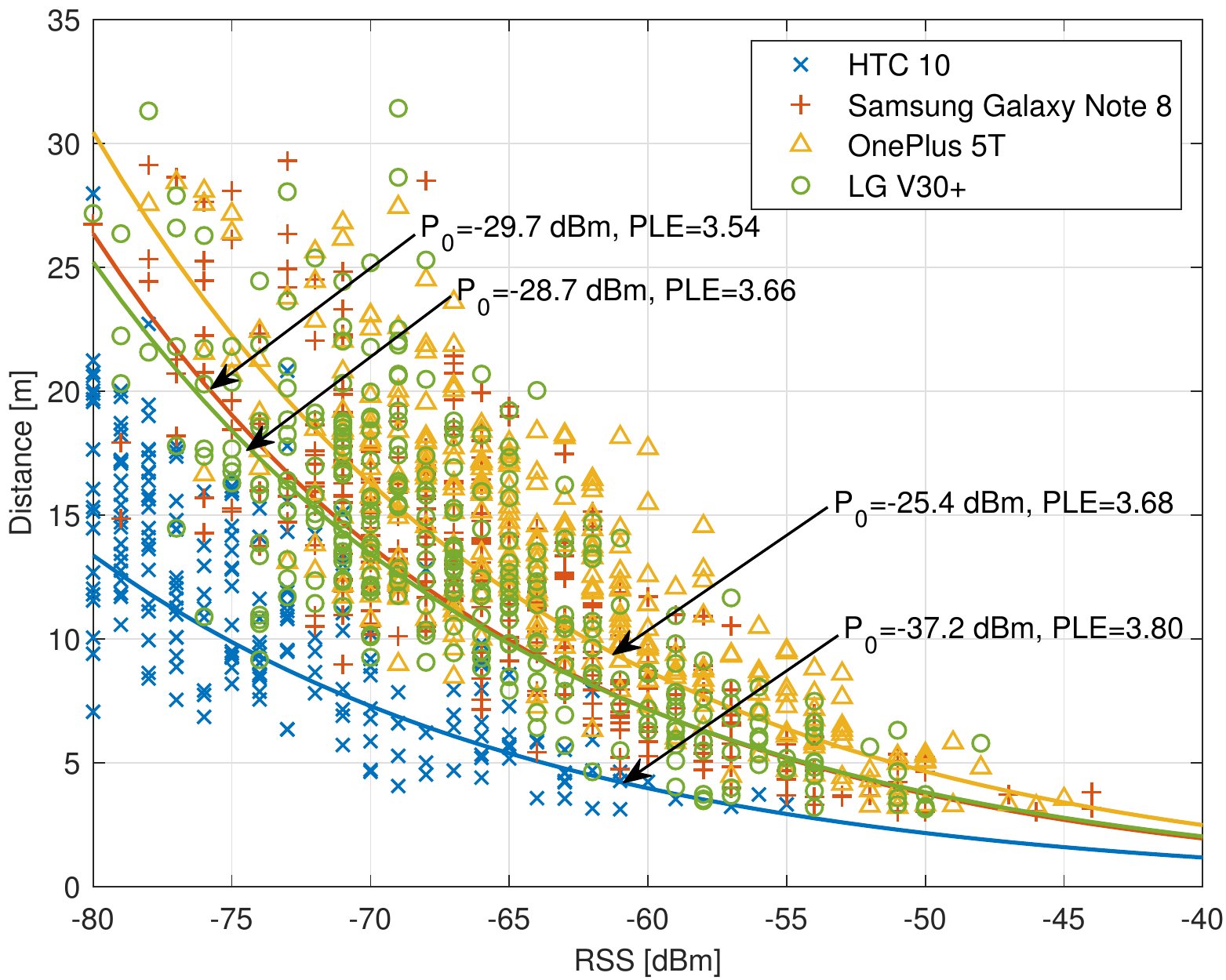}
    \caption{Relationship between actual distance and RSS measurements of 4 different mobile devices.}
    \label{fig_pathloss}
\end{figure}

The distance estimate from an AP can be expressed in the form of a parametric function as follows:
\begin{equation}
    \hat d = r(s; \Theta_{range}),
\end{equation}
where $s$ indicates the ranging source which is either an RSS or raw distance measurement using the FTM protocol, and $\Theta_{range}$ denotes the set of all parameters involved in the ranging procedure.

The pathloss model has been widely used for RSS-based ranging scenarios~\cite{Wang2003AnIW,5558044,5766644,7935650,8839041,8924707}.
This model expresses the RSS at distance $d$ from an AP as
\begin{equation} \label{3A1}
    P(d) = P_0 - 10\eta \log_{10} \frac{d}{d_0} + X,
\end{equation}
where $P_0$ is the RSS measured at a reference distance denoted by $d_0$, $\eta$ is the PLE, and $X$ represents a log-normal shadowing term that can be modeled as a zero mean random variable.
Using this relationship, the distance estimate from the AP is derived by
\begin{equation} \label{3A2}
    \hat{d}_{RSS} = r_{RSS}(P; \Theta_{RSS}) = d_0 10^{\frac{P_0 - P}{10 \eta}},
\end{equation}
where $P$ is current RSS measurement from the AP and $\Theta_{RSS} = \{P_0, \eta\}$ is the set of trainable parameters that should be selected carefully depending on the site and device.

Fig.~\ref{fig_pathloss} depicts the relationship between the measured RSS and the actual distance from APs, obtained using four different mobile devices. The details of the experiment are presented in Section~\ref{sec_exp}.
Although these data were collected from the same site along the same test path, measured RSS patterns widely vary from device to device.
The figure also indicates that the optimal ranging parameters, selected to have the minimum normalized mean squared error (NMSE) between the actual and estimated distances, are different for each device.

\begin{figure}
\centering
\captionsetup{farskip=0pt}
\subfloat[]{\includegraphics[width=0.22\textwidth]{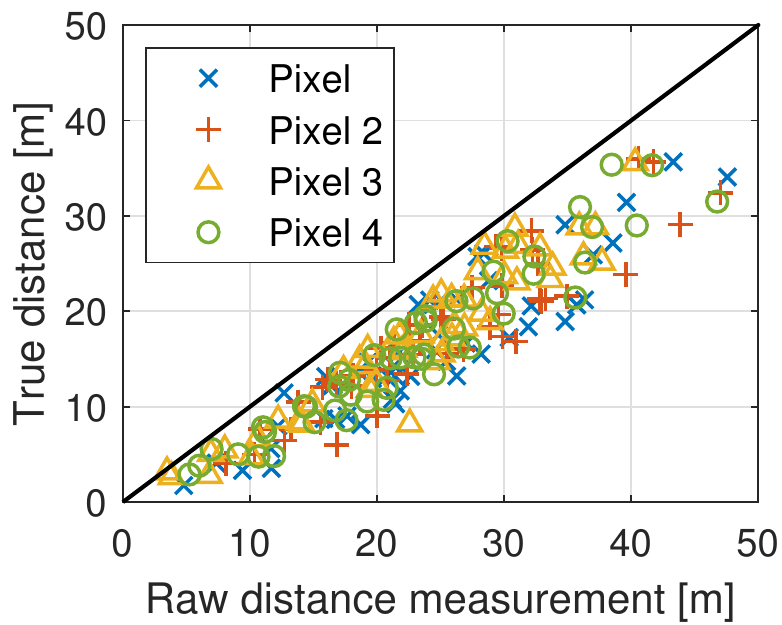}}\hfil\hfil
\subfloat[]{\includegraphics[width=0.22\textwidth]{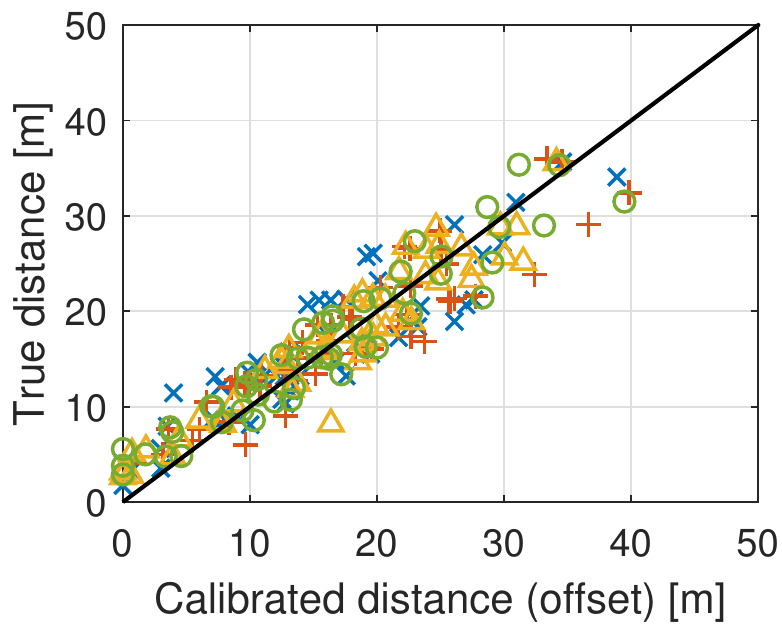}}\\
\subfloat[]{\includegraphics[width=0.22\textwidth]{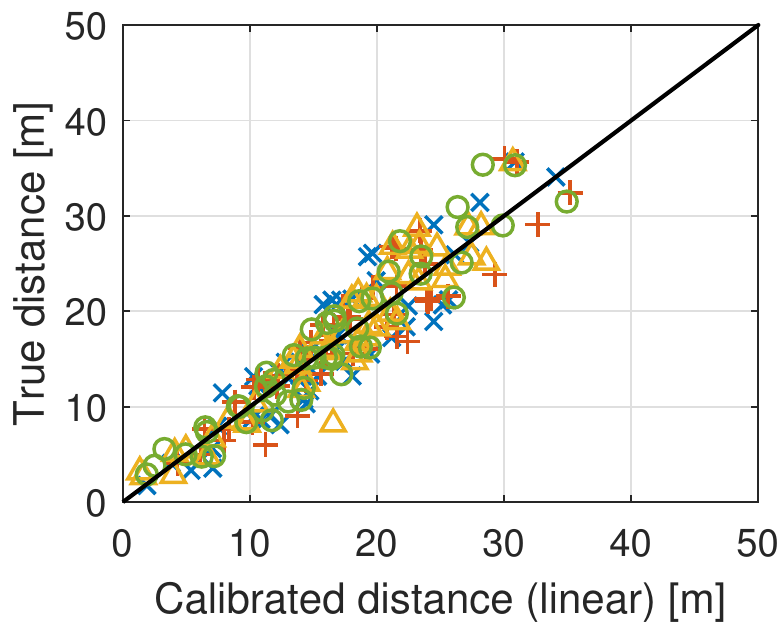}}\hfil\hfil
\subfloat[]{\includegraphics[width=0.22\textwidth]{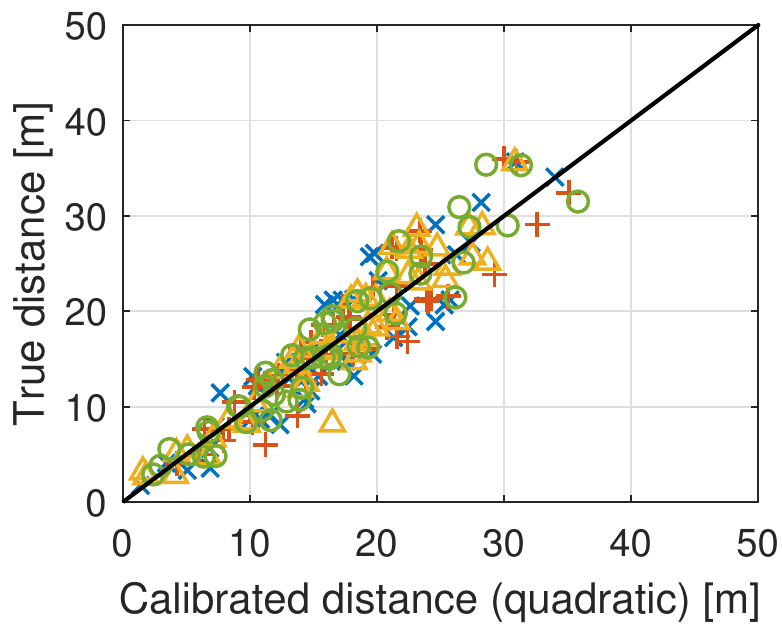}}
\caption{Relationship between the true distance and the following: (a) raw distance measurement from the FTM protocol, (b) calibrated distance with an offset ($m=1, c_1=1$), (c) calibrated distance with a linear polynomial ($m=1$), and (d) calibrated distance with a quadratic polynomial ($m=2$). The data were obtained using Google Pixel devices running on Android~10.}
\label{fig_ftm_scatter}
\end{figure}

If both AP and device support the FTM protocol, potentially more accurate ranging results can be obtained.
A well-known challenge is that distortions occur in the raw distance measurements obtained from the FTM protocol, which should be calibrated in advance to obtain accurate distances~\cite{ibrahim_mobicom_18,8839041,8911824,8911751,8924707}.
In this paper, we use a polynomial-based calibration curve to correct the distortion, where the calibrated distance from the AP is expressed by
\begin{equation} \label{3A3}
    \hat d_{RTT} = r_{RTT}(D; \Theta_{RTT}) = \max\left(\sum_{i=0}^{m} c_i D^i, 0\right).
\end{equation}
In this equation, $D$ indicates the raw distance measurement reported from the FTM protocol, and $m$ represents the degree of the calibration polynomial whose $i$-th order coefficient is denoted by $c_i$.
Furthermore, the $\max(\cdot)$ operator is applied to ensure that the calibration result becomes always non-negative.
For RTT-based ranging scenarios, the set of trainable parameters is given as $\Theta_{RTT} = \{c_0, c_1, ..., c_m\}$.

Fig.~\ref{fig_ftm_scatter} illustrates the relationship between the actual distance and various calibration results.
Fig.~\ref{fig_ftm_scatter}(a) indicates that the raw distance measurements from the FTM protocol of every device are biased from the black solid line, which represents the perfect distance estimation.
We first consider the simplest calibration method, which adds an offset to the raw distance measurement. In this case, $m=1$, $c_1=1$, and only $c_0$ is trainable.
The effect of an offset is illustrated in Fig.~\ref{fig_ftm_scatter}(b), where every point in Fig.~\ref{fig_ftm_scatter}(a) is shifted vertically toward the perfect line.
If a linear polynomial is applied (i.e., $m=1$), the slope of the points can be adjusted as shown in Fig.~\ref{fig_ftm_scatter}(c) depending on the value of $c_1$. In addition, a quadratic polynomial (i.e., $m=2$) can even correct the non-linear relationship of the points with $c_2$ as shown in Fig.~\ref{fig_ftm_scatter}(d).
The optimal parameters in the calibration polynomial are selected to minimize the mean squared error (MSE) between the actual and calibrated distance.

\subsection{Pedestrian Dead Reckoning Model}

The movement of the device on the positioning plane can be obtained from local accelerometer readings converted to the GCS.
For this, the orientation of the device in the GCS must be estimated.
In this paper, we represent the orientation of the device using a unit quaternion, which indicates that the device is rotated along an axis in the GCS.
A unit quaternion is expressed as a vector in $\mathbb{R}^{4\times 1}$ as follows:
\begin{align} \label{3B1}
\mathbf{q} = & [q_1, q_2, q_3, q_4]^T\nonumber\\
= & \left[e_x \sin\frac{\Phi}{2}, e_y \sin\frac{\Phi}{2}, e_z \sin\frac{\Phi}{2}, \cos\frac{\Phi}{2}\right]^T,
\end{align}
where $\mathbf e = [e_x, e_y, e_z]^T$ is a unit vector in the GCS indicating the rotation axis and $\Phi$ is the rotation angle.
In general, the quaternion is obtained using a Kalman filter (KF) with local sensor readings~\cite{Valenti2015,Feng2017,Bernal-Polo2019}.
Because orientation estimation is a widely researched topic, we will not discuss the details.

The relationship between two coordinate systems is also represented using a matrix in $\mathbb{R}^{3\times 3}$, called the direction cosine matrix (DCM). The $(i, j)$-th element of the DCM is defined by the cosine of angle between the $i$-th axis of the LCS and the $j$-th axis of the GCS ($i, j = 1,2,3$).
According to~\cite{Diebel06}, the DCM from a given unit quaternion $\mathbf{q}$ is derived as
\begin{equation} \label{3B2}
    \mathbf R = 
    \begin{bmatrix}
       1 - 2q_2^2 - 2q_3^2 & 2(q_1 q_2 + q_3 q_4) & 2(q_1 q_3 - q_2 q_4)\\
        2(q_1 q_2 - q_3 q_4) & 1 - 2q_1^2 - 2q_3^2 & 2(q_2 q_3 + q_1 q_4)\\
        2(q_1 q_3 + q_2 q_4) & 2(q_2 q_3 - q_1 q_4) & 1 - 2q_1^2 - 2q_2^2
    \end{bmatrix}.
\end{equation}
Using the DCM, the local accelerometer readings can be converted into the GCS as
\begin{equation} \label{3B3}
    \mathbf a = \mathbf R^T {\mathbf a}',
\end{equation}
where $\mathbf a = [a_x, a_y, a_z]^T$ and ${\mathbf a}' = [{a}_{x'}, {a}_{y'}, {a}_{z'}]^T$ indicate acceleration vectors on the GCS and the LCS, respectively\footnote{The DCM defined in this paper describes the orientation of the LCS relative to the GCS. We can define the DCM vice versa. For instance, the rotation matrix used in the Android application programming interface (API) indicates the DCM that describes the orientation of GCS relative to the LCS.}.

\begin{figure}
\captionsetup{farskip=0pt}
\centering
\subfloat[]{ \includegraphics[width=0.42\textwidth]{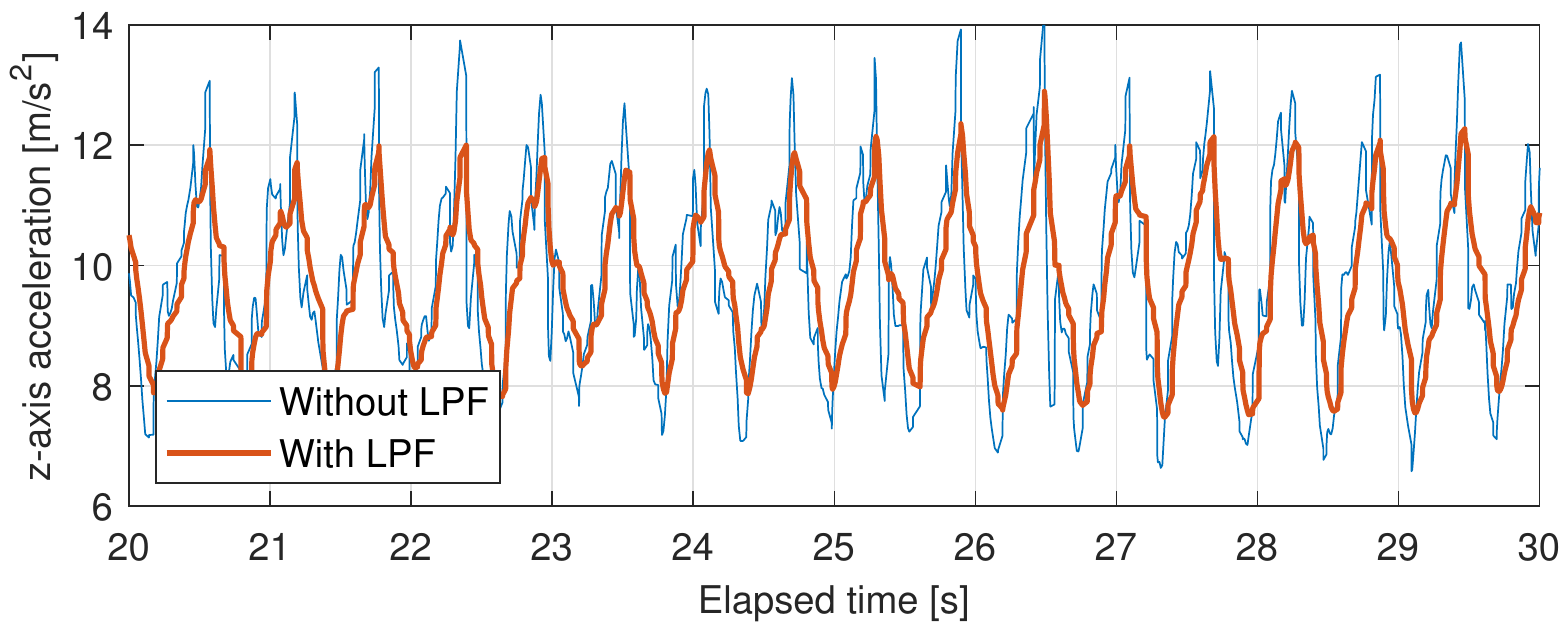}}\\
\subfloat[]{\includegraphics[width=0.42\textwidth]{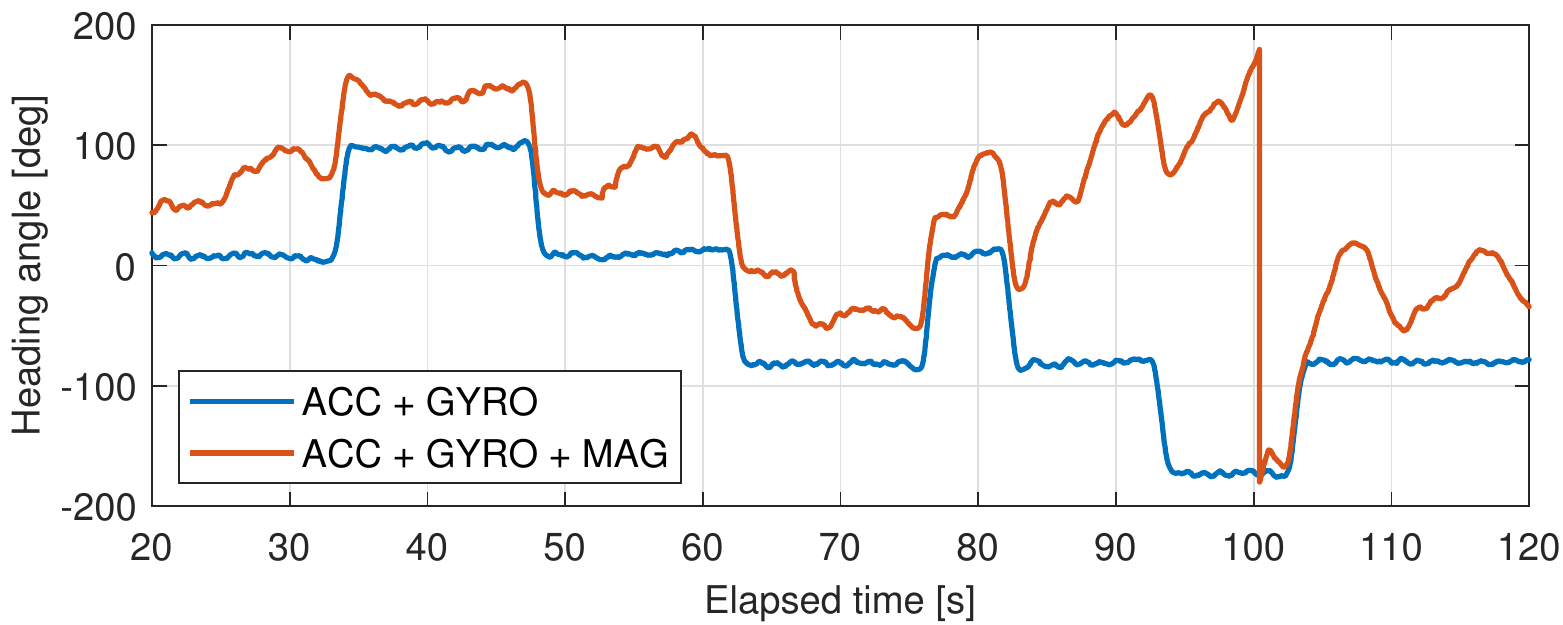}}
\caption{Examples of the PDR technique: (a) z-axis acceleration of the device on the GCS, and (b) heading estimation with and without the magnetometer. The actual heading was changed by +90 or -90 degrees in this experiment because users turned only left and right.}
\label{fig_pdr}
\end{figure}

The z-axis acceleration in the GCS is primarily used to detect the number of steps by capturing periodic up and down movement patterns of the human body, generated when the user is walking.
Moreover, a low pass filter (LPF) is applied to facilitate the step detection routine.
Fig.~\ref{fig_pdr}(a) illustrates an example of z-axis acceleration on the GCS and its LPF output.
A peak of the LPF output followed by a valley is considered as a step of the user.
In addition, the step length for each detected step can be calculated as~\cite{5646888}
\begin{equation} \label{3B4}
    l = \alpha \beta = \alpha (a_{z, max} - a_{z, min})^{\frac{1}{4}},
\end{equation}
where $\beta$ is a given value with the measured peak and valley acceleration, denoted by $a_{z,max}$ and $a_{z,min}$, respectively.
However, $\alpha$ is a step length coefficient that can vary from device to device or from user to user.

To estimate the heading direction, we reformulate the DCM using the rotation angles $\varphi$, $\psi$, and $\phi$ along the x-, y-, and z-axes on the GCS, respectively. These angles are also known as pitch, roll, and yaw angles.
According to the yaw-pitch-roll rotation sequence~\cite{Diebel06}, the DCM matrix is computed as
\begin{align} \label{3B5}
    \mathbf{R} = \mathbf{R}_y(\psi)\mathbf{R}_x(\varphi)\mathbf{R}_z(\phi) = 
    \begin{bmatrix}
        \cos\psi & 0 & -\sin\psi\\
        0 & 1 & 0 \\
        \sin\psi & 0 & \cos\psi
    \end{bmatrix}&\nonumber\\
    \begin{bmatrix}
        1 & 0 & 0\\
        0 & \cos\varphi & \sin\varphi \\
        0 & -\sin\varphi & \cos\varphi
    \end{bmatrix}
    \begin{bmatrix}
        \cos\phi & \sin\phi & 0\\
        -\sin\phi & \cos\phi & 0 \\
        0 & 0 & 1
    \end{bmatrix}&.
\end{align}
In this rotation sequence, the yaw angle represents the heading angle of the device relative to the GCS.
Because the $(2, 1)$- and $(2, 2)$-th elements of the above matrix multiplication are given by $[\mathbf{R}]_{(2, 1)} = -\cos\varphi\sin\phi$ and $[\mathbf{R}]_{(2, 2)}=\cos\varphi\cos\phi$, respectively, the heading angle can be obtained by comparing equations (\ref{3B2}) and (\ref{3B5}) as
\begin{equation} \label{3B6}
    \phi = -\arctan\frac{[\mathbf{R}]_{(2, 1)}}{[\mathbf{R}]_{(2, 2)}} = -\arctan \frac{2(q_1q_2 - q_3q_4)}{1-2q_1^2-2q_3^2}. 
\end{equation}

Fig.~\ref{fig_pdr}(b) shows an example of a heading estimation, where ACC, GYRO, and MAG in the legend stand for accelerometer, gyroscope, and magnetometer, respectively.
According to experimental results, the estimated heading angle with the magnetometer widely fluctuates even when the user moves along a straight path because of the distortion of the magnetic field in an indoor environment.
Hence, the magnetometer was not used in this study.
Without the magnetometer, the heading angle from equation (\ref{3B6}) is provided relative to an arbitrary reference direction on the positioning plane, instead of true x- and y-axes in the GCS.
To address this, we denote $\phi_{ref}$ as the reference heading direction, and thus, the correct heading angle of the device relative to the GCS is expressed by $\phi_{ref}$ + $\phi$ with $\phi$ obtained from equation (\ref{3B6}).

With the estimated step length and heading angle from sensor readings, the position of the device on the x-y plane of the GCS can be updated whenever each step is detected.
The estimated position of the device after $k \geq 1$ steps is given by
\begin{equation} \label{3B7}
\hat{\mathbf{p}}^{(k)}_{PDR} = \hat{\mathbf{p}}^{(k-1)}_{PDR} + \alpha \beta^{(k)} \mathbf{g}(\phi_{ref} + \phi^{(k)}),
\end{equation}
where $\beta^{(k)}$ is obtained using equation (\ref{3B4}) with the peak and valley acceleration of the $k$-th step, and $\phi^{(k)}$ is the heading angle when the $k$-th step is detected.
In addition, $\mathbf{g}(\phi) = [-\sin\phi, \cos\phi]^T$ indicates that the moving direction of the device on the GCS corresponds to the true heading angle $\phi$.
The estimated trajectory based on equation~(\ref{3B7}) clearly widely varies depending on many parameters, including the starting position of the device, denoted by $\hat{\mathbf p}_{PDR}^{(0)} = [x_0, y_0]^T$.
Therefore, we denote the set of parameters related to the PDR model as $\Theta_{PDR} = \{x_0, y_0, \phi_{ref}, \alpha\}$.



\section{Positioning with Online Calibration}

\begin{figure}
    \centering
    \includegraphics[width=0.45\textwidth]{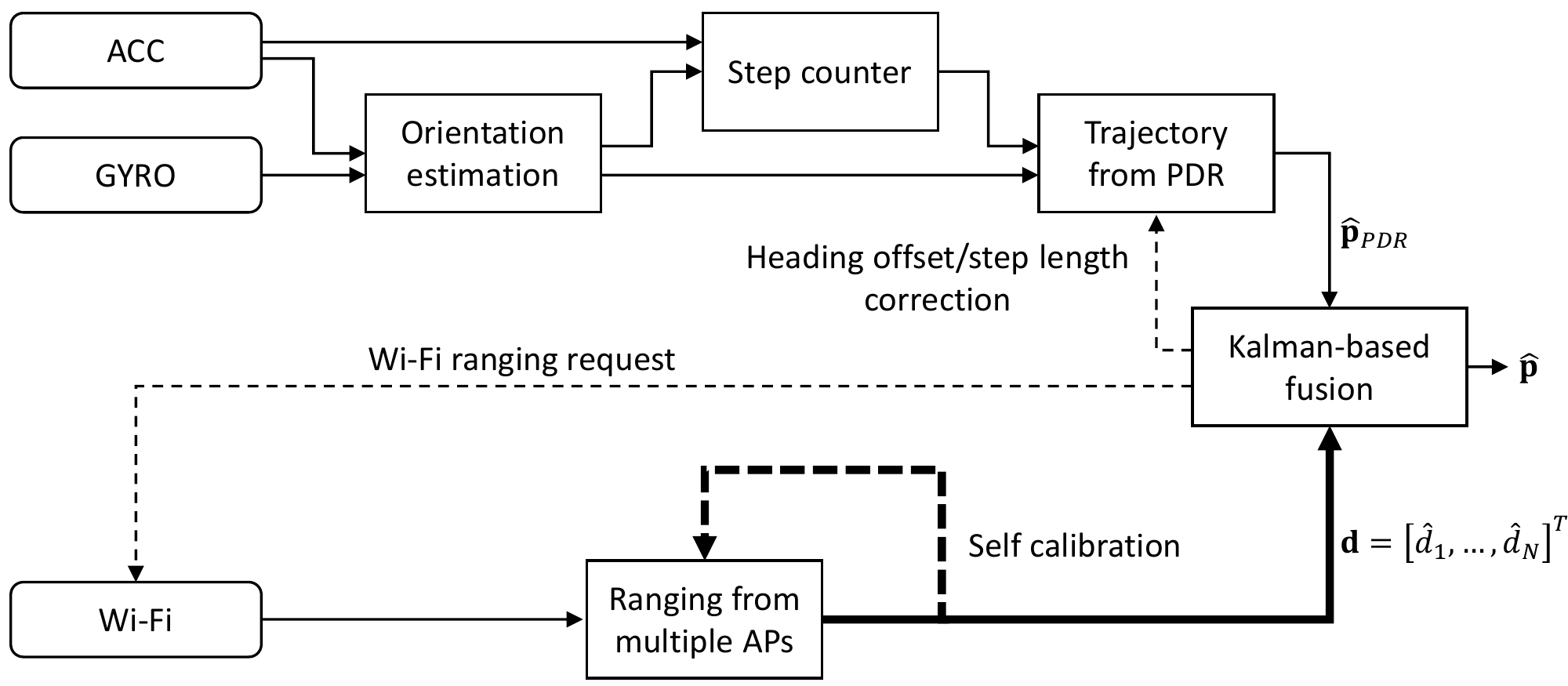}
    \caption{Overview of the proposed online calibration framework.}
    \label{fig_overview}
\end{figure}

Fig.~\ref{fig_overview} illustrates the proposed online calibration framework consisting of Wi-Fi ranging and PDR modules.
After a few steps are detected in the PDR module, the proposed method performs an initial calibration procedure to simultaneously estimate all parameters in the system.
After the initial calibration, a KF is applied to estimate the position of the device by combining the step detection and Wi-Fi ranging results.
At the same time, the ranging module periodically performs a self-calibration routine to optimize ranging parameters.
On the other hand, the parameters in the PDR module are updated as part of the KF state.
In addition, the proposed method invokes Wi-Fi ranging procedures depending on the state estimation results from the KF.
In this paper, we proceed with a KF time step when a step is detected in the PDR module. Therefore, time step $k$ indicates that $k$ steps were detected in the PDR module.

\subsection{Initial Calibration}
\label{sec_init_cal}

Initial calibration is performed by combining outputs from both Wi-Fi ranging and PDR modules.
For this, we perform a Wi-Fi ranging procedure whenever each step is detected until the first $B$ steps are detected, where $B$ is a constant (e.g., $B=8$).
After the last ranging procedure, the initial calibration procedure is executed to minimize the following cost function:
\begin{equation} \label{4A1}
    J({\Theta}) = \sum_{k=1}^{B}\sum_{n=1}^N \epsilon_{n, k}^2 =  \sum_{k=1}^{B}\sum_{n=1}^{N} \left(\lVert\hat{\mathbf p}_{PDR}^{(k)} -\mathbf{p}_n^{(k)}\rVert - \hat{d}_n^{(k)}\right)^2,
\end{equation}
where ${\Theta} = \Theta_{range} \cup \Theta_{PDR}$ represents the set of all parameters in the system and $N$ is the number of APs selected for positioning at each time step. 
Moreover, $\mathbf{p}_n^{(k)}$ represents the position of the $n$-th selected AP at time step $k$, and $\hat{d}^{(k)}_n$ is the distance estimate from this AP using either equation (\ref{3A2}) or (\ref{3A3}) depending on the ranging source. Accordingly, $\Theta_{range}$ indicates either $\Theta_{RSS}$ or $\Theta_{RTT}$.

The optimal parameters that minimize the cost function can be obtained using a simple gradient descent method. For this, the derivative of the cost function with respect to a parameter $\theta \in \Theta$ is computed as
\begin{equation} \label{4A2}
    \frac{\partial J({\Theta})}{\partial \theta} = \sum_{k=1}^B \sum_{n=1}^N 2\epsilon_{n, k}\left(\frac{(\hat{\mathbf p}_{PDR}^{(k)} - \mathbf p_n^{(k)})^T \frac{\partial \hat{\mathbf p}_{PDR}^{(k)}}{\partial \theta}}{\lVert\hat{\mathbf p}_{PDR}^{(k)}- \mathbf p_n^{(k)}\rVert} - \frac{\partial \hat d_n^{(k)}}{\partial \theta}\right).
\end{equation}
According to the derivative obtained from the above equation, each parameter is iteratively updated in the direction of minimizing the cost function as 
\begin{equation} \label{4A3}
    \theta \leftarrow \theta - \lambda \frac{\partial J(\Theta)}{\partial \theta},~\theta \in\Theta,
\end{equation}
where $\lambda$ represents the learning rate.
The optimization iteration can be executed up to a predefined maximum number of iterations, or stop early if the parameter update does not improve the cost function significantly.

\subsection{Self-Calibration of Wi-Fi Ranging Module}

As Wi-Fi ranging procedures are executed multiple times during the operation of the positioning application, the amount of training data for optimizing ranging parameters accumulate.
In this paper, the cost function for optimizing parameters using all accumulated Wi-Fi ranging results is simply defined as the sum of the cost of each individual result.
Therefore, we first focus on the cost function design for a Wi-Fi ranging result at a specific time step. For this reason, the time step index is omitted at the beginning of this subsection.

The cost function for ranging is defined as the square sum of Wi-Fi ranging errors as follows:
\begin{equation} \label{4B1}
    C(\mathbf{p}; \Theta_{range}) = \sum_{n=1}^N \epsilon_n^2 = \sum_{n=1}^N \left(\lVert\mathbf{p} - \mathbf{p}_n\rVert -  \hat{d}_n\right)^2,
\end{equation}
where $\mathbf{p}$ is the position of the device and $\hat{d}_n$ represents the estimated distance from the $n$-th AP.
Because this cost function also depends on the position of the device, we consider the best achievable cost with current estimates of ranging parameters as follows:
\begin{equation} \label{4B2}
    \tilde{C}(\Theta_{range}) = C(\mathbf{p}^*; \Theta_{range}),
\end{equation}
where ${\mathbf p}^* = \argmin_{\mathbf p} C(\mathbf p; \Theta_{range})$ represents an optimal position that minimizes the cost function.
The above best achievable cost function always produces non-negative values and becomes 0 if every distance estimate is perfect.

Similar to the initial calibration procedure, we can iteratively update each ranging parameter in the direction of optimizing the best achievable cost.
For this, the derivative of the cost function in equation (\ref{4B2}) with respect to a parameter $\theta \in \Theta_{range}$ is derived as
\begin{equation} \label{4B3}
    \frac{\partial \tilde{C}(\Theta_{range})}{\partial \theta} = \sum_{n=1}^N 2\epsilon_n\left(\frac{({\mathbf p}^* - \mathbf p_n)^T \frac{\partial {\mathbf p}^*}{\partial \theta}}{\lVert{\mathbf p}^*-\mathbf p_n\rVert} - \frac{\partial \hat{d}_n}{\partial \theta}\right).
\end{equation}
Because $\frac{\partial {\mathbf p}^*}{\partial \theta}$ should be computed to obtain the above derivative, we use the linear least-square (LS) method to approximate $\mathbf{p}^*$ as follows~\cite{1275684, paula_sen_2011}:
\begin{equation} \label{4B4}
    {\mathbf p}^* = (\mathbf{A}^T \mathbf{A})^{-1} \mathbf{A}^T \mathbf b,
\end{equation}
where $\mathbf A\in \mathbb{R}^{(N-1)\times 2}$ and $\mathbf b \in \mathbb{R}^{(N-1) \times 1}$ are defined by
\begin{align} \label{4B5}
    \mathbf A 
    = 2\begin{bmatrix}
    (\mathbf p_2 - \mathbf p_1)^T\\[1pt]
    \vdots\\[1pt]
    (\mathbf p_N - \mathbf p_1)^T\\
    \end{bmatrix},
    \mathbf b
    = \begin{bmatrix}
    \lVert\mathbf p_2\rVert^2 - \lVert\mathbf p_1\rVert^2 - \hat{d}_2^2 + \hat{d}_1^2\\[1pt]
    \vdots\\[1pt]
    \lVert\mathbf p_N\rVert^2 - \lVert\mathbf p_1\rVert^2 - \hat{d}_N^2 + \hat{d}_1^2
    \end{bmatrix}.
\end{align}
Using this relationship, the derivative of $\mathbf p^*$ with respect to $\theta$ is simply computed by
\begin{equation} \label{4B6}
\frac{\partial {\mathbf p}^*}{\partial \theta}
= (\mathbf A^T \mathbf A)^{-1}\mathbf A^T 
\begin{bmatrix}
    -2\hat{d}_2 \frac{\partial \hat{d}_2}{\partial \theta} + 2\hat{d}_1 \frac{\partial \hat{d}_1}{\partial \theta}\\[1pt]
    \vdots\\[1pt]
    -2\hat{d}_N \frac{\partial \hat{d}_N}{\partial \theta} + 2\hat{d}_1 \frac{\partial \hat{d}_1}{\partial \theta}
\end{bmatrix},
\end{equation}
and the derivative in equation (\ref{4B3}) is obtained accordingly.

If multiple Wi-Fi ranging results are available, the same process is applied to each ranging result.
Now, we assume that $M$ ranging results are accumulated just before executing the self-calibration routine and denote $\tilde{C}_i(\Theta_{range})$ as the best achievable cost related to the $i$-th ranging result.
In this case, each ranging parameter is updated by
\begin{equation}
    \theta \leftarrow \theta - \tilde{\lambda}\sum_{i=1}^M \frac{\partial \Tilde{C}_i(\Theta_{range})}{\partial \theta},~\theta\in\Theta_{range},
\end{equation}
where $\tilde{\lambda}$ is the learning rate for optimizing parameters in the ranging module.

\subsection{KF-Based Positioning and Calibration of the PDR Module}

When the initial calibration is complete, we can obtain coarse estimates of device's position and parameters in the PDR module.
In this subsection, we design a KF to obtain more accurate estimates using a series of measurements over time.
In particular, we use an extended KF (EKF) because both state transition and measurement models are non-linear to the KF state, which is defined by a vector in $\mathbb{R}^{4\times 1}$ as follows:
\begin{equation} \label{4C1}
    \mathbf z = [x, y, \phi_{ref}, \alpha]^T = [\mathbf p^T, \phi_{ref}, \alpha]^T.
\end{equation}
At each time step of the KF procedure, we predict the state from the previously estimated state using the output from the PDR module and correct the predicted state using Wi-Fi ranging results.
We denote $\hat{\mathbf z}^{(i|j)}$ as the state estimate at time step $i$ using every measurement until time step $j$ ($i\geq j$). 
The state estimate is considered as a multivariate Gaussian distributed vector whose covariance matrix is defined by
\begin{equation} \label{4C2}
    \mathbf{P}^{(i|j)} = E\left[(\hat{\mathbf z}^{(i|j)} - \boldsymbol{\mu}^{(i|j)})(\hat{\mathbf z}^{(i|j)} - \boldsymbol{\mu}^{(i|j)})^T\right],
\end{equation}
where $\boldsymbol{\mu}^{(i|j)} = E[\hat{\mathbf{z}}^{(i|j)}]$ represents the mean vector of $\hat{\mathbf z}^{(i|j)}$.

\emph{1) State Initialization}: The KF state is initialized at time step~$B$ using the initial calibration results.
For instance, the initial state of x, y coordinates is given by $[\hat{x}^{(B|B)}, \hat{y}^{(B|B)}]^T = \hat{\mathbf p}_{PDR}^{(B)}$.
In addition, the covariance matrix corresponding to the initial state is simply given as a diagonal matrix as
\begin{equation} \label{4C3}
    \mathbf P^{(B|B)} = \mbox{diag}(\sigma_x^2, \sigma_y^2, \sigma_\phi^2, \sigma_\alpha^2),
\end{equation}
where each diagonal element represents the variance of initial estimate of each element in the state.

\emph{2) State Prediction with PDR Module Outputs}: Whenever the PDR module detects a new step, the KF predicts the current state from the last state estimate.
The state transition model for time step $k>B$ is expressed by
\begin{equation} \label{4C4}
    \hat{\mathbf z}^{(k|k-1)} = \mathbf f\left(\hat{\mathbf{z}}^{(k-1|k-1)}, \mathbf u^{(k)}(\hat{\mathbf{z}}^{(k-1|k-1)})\right) + \mathbf{w}^{(k)},
\end{equation}
where $\mathbf u^{(k)}(\cdot)$ represents the position change due to the $k$-th step, which is obtained from equation~(\ref{3B7}).
Because the position change depends on the parameters in the PDR module, $\mathbf u^{(k)}(\cdot)$ is expressed as a function of the state as follows:
\begin{equation}
    \mathbf{u}^{(k)}(\mathbf z) = \alpha \beta^{(k)}\mathbf{g}(\phi_{ref} + \phi^{(k)}).
\end{equation}
In addition, $\mathbf w^{(k)} \in \mathbb{R}^{4\times 1}$ indicates the state transition error, which is assumed to be a zero mean multivariate Gaussian vector with the covariance matrix of $\mathbf Q^{(k)} = E[\mathbf w^{(k)}(\mathbf w^{(k)})^T]$.
The transition of each element in the state is given by
\begin{gather}
    \hat{\mathbf p}^{(k|k-1)} = \hat{\mathbf{p}}^{(k-1|k-1)} + \mathbf u^{(k)}(\hat{\mathbf z}^{(k-1|k-1)}), \nonumber\\
    \hat{\phi}_{ref}^{(k|k-1)} = \hat{\phi}_{ref}^{(k-1|k-1)},~\hat{\alpha}^{(k|k-1)} = \hat{\alpha}^{(k-1|k-1)}.
    \label{4C5}
\end{gather}

To update the covariance matrix according to the state transition results, the Jacobian of the state transition function $\mathbf{f}(\cdot)$ should be computed as
\begin{align} \label{4C6}
\mathbf{F}^{(k)} = \evalat[\Big]{\frac{\partial \mathbf f (\mathbf z, \mathbf{u}^{(k)}(\mathbf z))}{\partial {\mathbf z}}}{\mathbf z = \hat{\mathbf z}^{(k-1|k-1)}}
= \begin{bmatrix}
\mathbf{I}_{2\times 2} & [\mathbf{v}_\phi^{(k)}, \mathbf{v}_\alpha^{(k)}] \\
\mathbf{0}_{2\times 2} & \mathbf{I}_{2\times 2}
\end{bmatrix},
\end{align}
where $\mathbf{v}_{\phi}^{(k)}, \mathbf{v}_{\alpha}^{(k)} \in \mathbb{R}^{2\times1}$ are respectively derived as
\begin{gather}
\evalat[\Big]{\frac{\partial \mathbf u^{(k)}(\mathbf z)}{\partial \phi_{ref}}}{\mathbf z = \hat{\mathbf z}^{(k-1|k-1)}}
= \hat{\alpha}^{(k-1|k-1)} \beta^{(k)}\tilde{\mathbf{g}}(\hat{\phi}_{ref}^{(k-1|k-1)} + \phi^{(k)}),\nonumber\\
    \evalat[\Big]{\frac{\partial \mathbf{u}^{(k)}(\mathbf z)}{\partial \alpha}}{\mathbf z = \hat{\mathbf z}^{(k-1|k-1)}} = \beta^{(k)} {\mathbf g}(\hat{\phi}_{ref}^{(k-1|k-1)}+\phi^{(k)}).
    \label{4C7}
\end{gather}
Here, $\tilde{\mathbf g}(\phi) = [-\cos\phi, -\sin\phi]^T$ represents the derivative of $\mathbf g(\phi)$ with respect to $\phi$.
With this result, the covariance matrix is updated as
\begin{equation} \label{4C8}
    \mathbf{P}^{(k|k-1)} = \mathbf F^{(k)} \mathbf{P}^{(k-1|k-1)} (\mathbf F^{(k)})^T + \mathbf{Q}^{(k)}.
\end{equation}

\emph{3) Condition for New Wi-Fi Ranging Request}: The main intuition to perform an irregular Wi-Fi ranging procedure is that the covariance matrix $\mathbf P^{(k|k-1)}$ contains the variance of each element in the predicted state, and if the variance of a particular element is sufficiently small enough, the current estimate of that element can be considered relatively accurate.
Therefore, if the variances of some parameter estimates are within an acceptable range, the Wi-Fi ranging procedure can be skipped.

Because the main interest in this paper is to achieve accurate positioning results, we only consider the variance of x and y coordinate estimates.
Therefore, we propose a simple condition for initiating a new Wi-Fi ranging request as follows:
\begin{equation}
   \sqrt{[\mathbf{P}^{(k|k-1)}]_{(1,1)}+[\mathbf{P}^{(k|k-1)}]_{(2,2)}} > \rho,
\end{equation}
where $\rho$ represents a threshold that can be chosen depending on the required positioning accuracy level of the application. For instance, if an application requires the best positioning accuracy, $\rho$ is selected as 0 to invoke the Wi-Fi ranging procedure as often as possible.

\emph{4) State Update with Wi-Fi Ranging Results}: If a Wi-Fi ranging procedure was performed at the current time step and estimated distances from nearby APs are available, we can correct the predicted state using these measurements.
We denote $\mathbf d^{(k)} = [\hat d_1^{(k)}, ..., \hat d_N^{(k)}]^T$ as the distance measurement vector from $N$ nearby APs at time step $k$, and it can be modeled as follows:
\begin{equation}
    \mathbf{d}^{(k)} = 
    \mathbf h(\mathbf z) + \boldsymbol{\upsilon}^{(k)} = 
    \begin{bmatrix}
        \lVert\mathbf p - \mathbf p_1^{(k)}\rVert \\[1pt]
        \vdots\\[1pt]
        \lVert\mathbf p - \mathbf p_N^{(k)}\rVert
    \end{bmatrix}
    + \boldsymbol{\upsilon}^{(k)},
\end{equation}
where $\boldsymbol{\upsilon}^{(k)} = [\upsilon_1^{(k)}, ..., \upsilon_N^{(k)}]^T$ is the distance measurement error vector with zero mean and the covariance matrix of $\boldsymbol{\Lambda}^{(k)} = E[\boldsymbol{\upsilon}^{(k)} (\boldsymbol{\upsilon}^{(k)})^T]$.
Note that if there are less than $N$ APs available in practice, the dimension of $\mathbf d^{(k)}$ is reduced to include only available APs.

With this measurement model, the innovation of the KF is defined as the difference between measured and expected distances as
\begin{equation}
    \boldsymbol{\zeta}^{(k)} = \mathbf d^{(k)} - \mathbf h(\hat{\mathbf z}^{(k|k-1)}).
\end{equation}
The covariance matrix of the innovation is derived as 
\begin{equation}
    \mathbf{S}^{(k)} = \mathbf{H}^{(k)}\mathbf{P}^{(k|k-1)} (\mathbf {H}^{(k)})^T + \boldsymbol{\Lambda}^{(k)},
\end{equation}
where $\mathbf {H}^{(k)}\in \mathbb{R}^{N\times 4}$ is a Jacobian matrix defined by
\begin{equation}
    \mathbf{H}^{(k)} = \evalat[\Big]{\frac{\partial \mathbf{h}(\mathbf z)}{\partial \mathbf z}}{\mathbf z = \hat{\mathbf z}^{(k|k-1)}}
    = \begin{bmatrix}
        \frac{(\hat{\mathbf p}^{(k|k-1)} - \mathbf p_1^{(k)})^T}{\lVert\hat{\mathbf p}^{(k|k-1)} - \mathbf p_1^{(k)}\rVert} & \mathbf 0_{1\times 2} \\[1pt]
        \vdots & \vdots\\[1pt]
        \frac{(\hat{\mathbf p}^{(k|k-1)} - \mathbf p_N^{(k)})^T}{\lVert\hat{\mathbf p}^{(k|k-1)} - \mathbf p_N^{(k)}\rVert} & \mathbf 0_{1\times 2} 
    \end{bmatrix}.
\end{equation}
Using above equations, the Kalman gain, updated state and its covariance matrix are derived as
\begin{gather}
    \mathbf K^{(k)} = \mathbf{P}^{(k|k-1)}(\mathbf H^{(k)})^T(\mathbf{S}^{(k)})^{-1},\nonumber\\
    \hat{\mathbf z}^{(k|k)} = \hat{\mathbf z}^{(k|k-1)} + \mathbf{K}^{(k)}\boldsymbol{\zeta}^{(k)},\nonumber\\
    \mathbf{P}^{(k|k)} = (\mathbf{I}_{4\times 4} - \mathbf{K}^{(k)} \mathbf{H}^{(k)})\mathbf{P}^{(k|k-1)}.
\end{gather}
Note that $\hat{\mathbf z}^{(k|k)}$ indicates the estimate of the current state using every available measurement.
If Wi-Fi ranging procedure is not performed at time step $k$, we simply define $\hat{\mathbf{z}}^{(k|k)} = \hat{\mathbf{z}}^{(k|k-1)}$ and $\mathbf{P}^{(k|k)} = \mathbf{P}^{(k|k-1)}$ obtained from equations (\ref{4C4}) and (\ref{4C8}), respectively.


\section{Experimental Results} \label{sec_exp}

\subsection{Experiment Setup}

\begin{figure}
    \centering
    \includegraphics[width=0.42\textwidth]{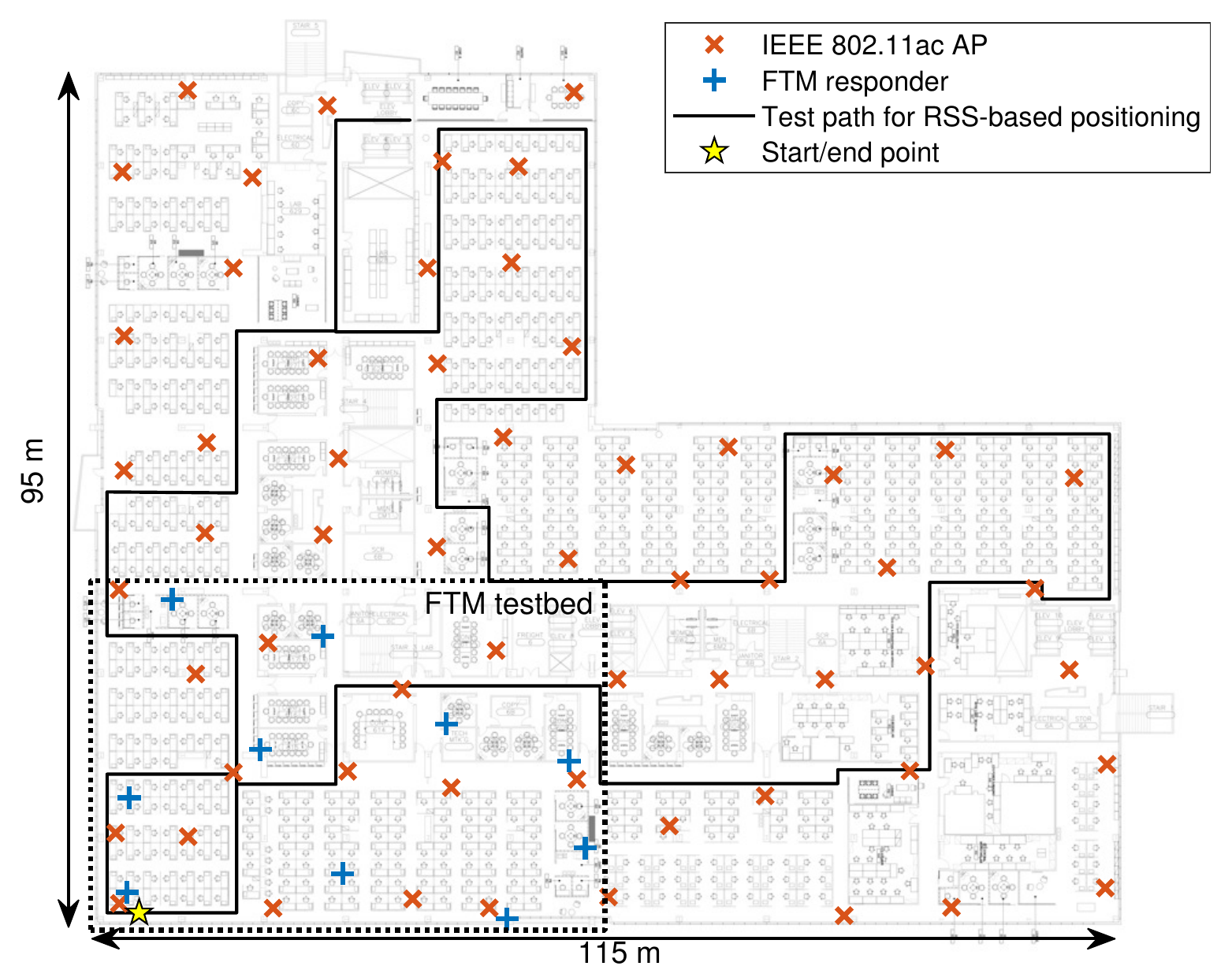}
    \caption{Floor plan of the experiment site, where 59 IEEE 802.11ac APs and 10 IEEE 802.11-2016 capable APs were installed.}
    \label{fig_map}
\end{figure}

\begin{table}
\caption{List of Android devices used in the experiments}
\label{table_dev}
\renewcommand{\arraystretch}{1.2}
\centering
\begin{tabular}{c|c|c|c}
	Device \# & Model name & Release date & Version\\
\hline
    1 & HTC 10 & 2016/04 & 8.0\\
	2 & Samsung Galaxy Note 8 & 2017/08 & 7.1.1 \\
	3 & OnePlus 5T & 2017/11 & 7.1.1\\
	4 & LG V30+ & 2018/03 & 8.0.0\\
\hline
	5 & Google Pixel & 2016/10 & 10 \\
	6 & Google Pixel 2 & 2017/10 & 10\\
	7 & Google Pixel 3 & 2018/10 & 10\\
	8 & Google Pixel 4 & 2019/10 & 10\\
\hline
\end{tabular}
\end{table}

Fig.~\ref{fig_map} depicts the floor plan of the experimental site, where the maximum height and width are 95 and 115~m, respectively. 
This site consists of various materials such as concrete, wood, metal walls, and various furniture.
In this site, 59 IEEE 802.11ac APs are installed on the ceiling to cover the entire area, and they broadcast beacon on both 2.4 and 5~GHz bands.
We only consider RSSs on a 2.4~GHz band for the RSS-based ranging scenarios.

Because the existing APs do not support the FTM protocol, we additionally installed 10 FTM responders (FTMRs) in a 56~m$\times$37~m test area to verify the positioning performance for RTT-based ranging scenarios.
The FTMRs are equipped with an Intel AC8260 Wi-Fi device and they respond to ranging requests from mobile devices (i.e., FTM initiators). 
Each FTMR operates on a 5~GHz band using Wi-Fi channel 40 or 48, and the bandwidth for ranging is configured to 40~MHz as default.
All FTMRs are installed on top of cubicles or in the conference room instead of on the ceiling.

The list of mobile devices used in this experiment is summarized in Table~\ref{table_dev}.
Among many devices we tested, only the Google Pixel devices running on Android 9 or above supported the FTM protocol.
For this reason, we evaluated the positioning performance of RTT-based ranging scenarios using the Google Pixel devices and the performance of RSS-based ranging scenarios using the other devices in the table.
A real-time Android application was implemented for the experiment, where the implementation details related to Wi-Fi ranging procedures were discussed in~\cite{8839041}.

The application scans only Wi-Fi channels 1, 6, and 11 on a 2.4~GHz band using the customized RSS scan method to quickly obtain RSS measurements~\cite{custom_scan}.
This enables us to obtain RSS measurement results every 500~ms for almost every device.
Similarly, the RTT-based ranging results also can be obtained less than 500~ms for the Google Pixel series.
For the PDR module, we collected the accelerometer and gyroscope readings using a sampling rate of 100~Hz.
Every operation, such as orientation estimation of the device, initial calibration, KF for state estimation, was implemented in the real-time application.
A demo video is available online~\cite{demo_video}.

\begin{figure}
    \centering
    \includegraphics[width=0.42\textwidth]{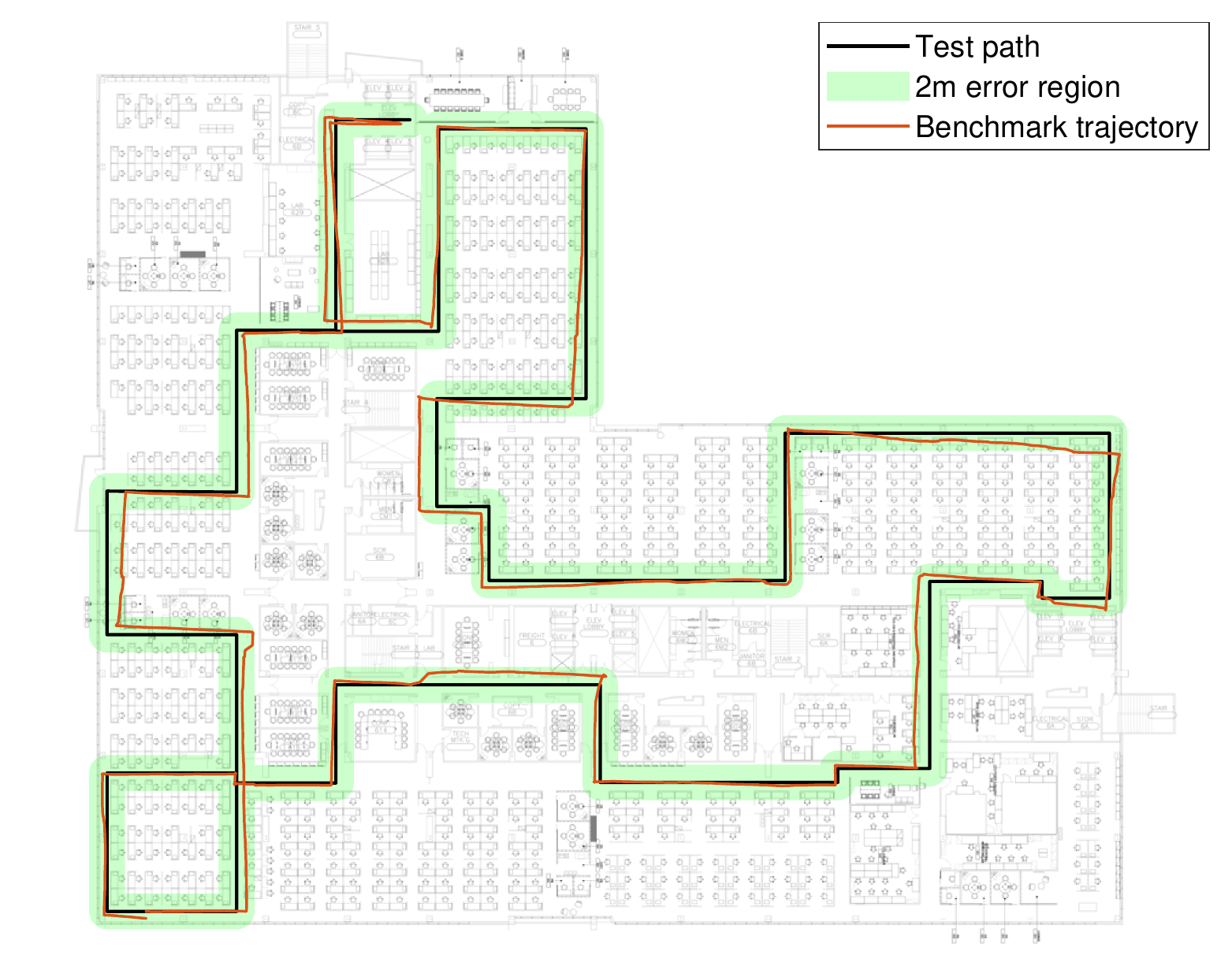}
    \caption{Estimated trajectory for the benchmark scenario using Samsung Galaxy Note~8.}
    \label{fig_rss_benchmark}
\end{figure}

\subsection{Positioning with RSS-Based Ranging and PDR}

For performance comparison, we first consider a benchmark scenario, where every parameter in the system is perfectly optimized to produce the best results for the test path shown in Fig.~\ref{fig_map}.
We remind that the set of ranging and PDR parameters are given by $\Theta_{RSS} = \{P_0, \eta\}$ and $\Theta_{PDR} = \{x_0, y_0, \phi_{ref}, \alpha\}$, respectively.
In the benchmark scenario, we performed the Wi-Fi ranging procedure as rapidly as possible (i.e., every 500~ms) to observe the best achievable positioning performance.
The ranging parameters were optimized by collecting RSS measurements along the test path.
Because the true coordinates of the device were also measured to evaluate the positioning performance, we could have actual distance corresponding to each RSS measurement.
Therefore, the optimal ranging parameters were selected to have the minimum NMSE between the estimated and actual distances.
If another test path is used to evaluate the performance, we can follow the exact same process to find optimal parameters that produce the best performance on that test path.

\begin{figure}
\captionsetup{farskip=0pt}%
\centering
\subfloat[]{\includegraphics[width=0.22\textwidth]{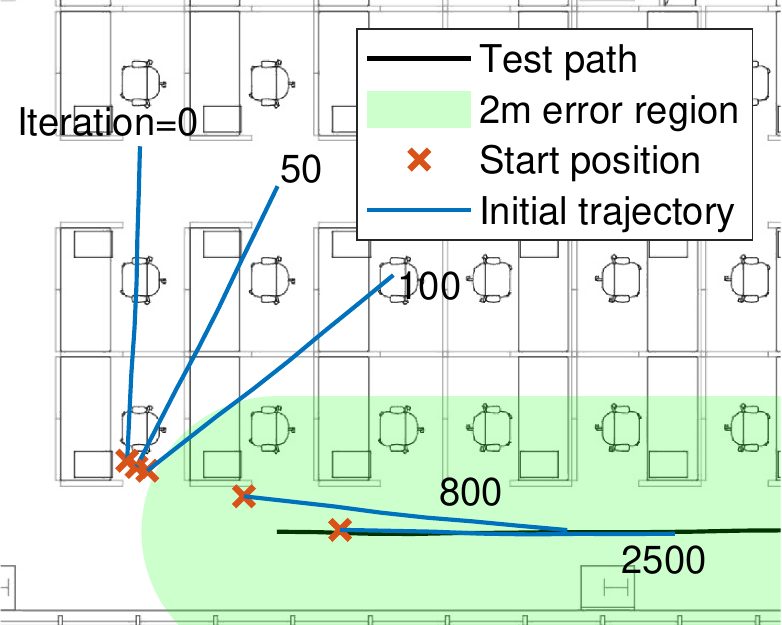}}\hfil\hfil\hfil
\subfloat[]{\includegraphics[width=0.22\textwidth]{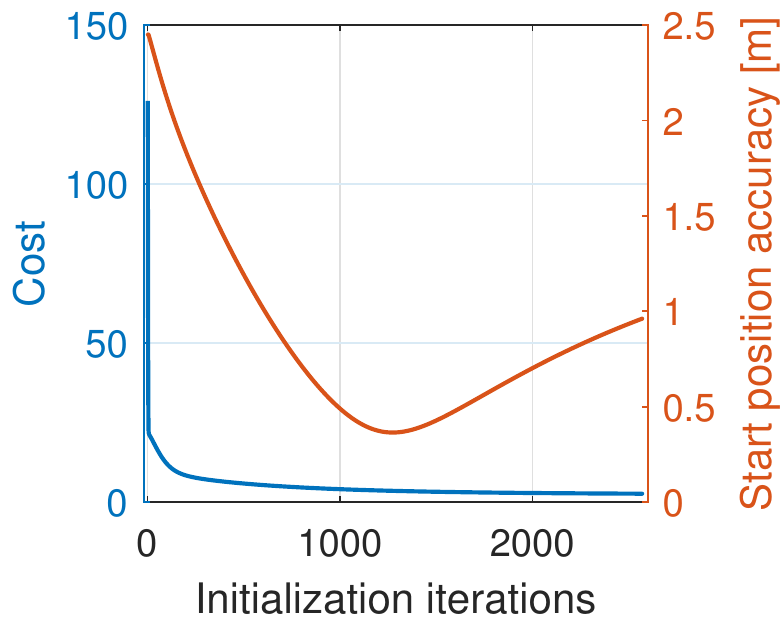}}\vfil
\subfloat[]{\includegraphics[width=0.22\textwidth]{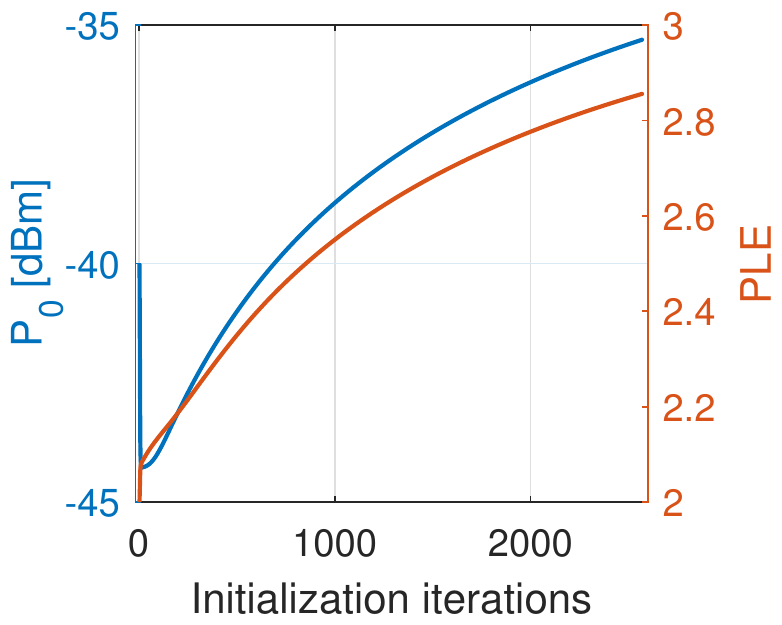}}\hfil\hfil\hfil
\subfloat[]{\includegraphics[width=0.22\textwidth]{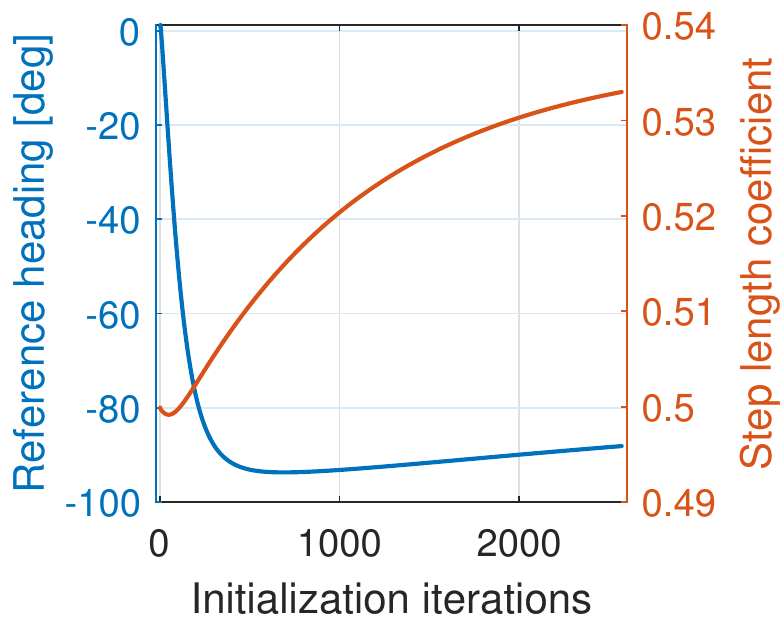}}
\caption{Example of initial calibration step: (a) estimated trajectory, (b) cost function and start position accuracy, (c) ranging parameters, and (d) PDR parameters.}
\label{fig_rss_init}
\end{figure}

\begin{table}
\caption{Benchmark Parameters and Online Calibration Results for the RSS-based Ranging Scenario}
\label{table_rss_parm}
\renewcommand{\arraystretch}{1.2}
\centering
\begin{tabular}{c|cccc}
 Device \# & 1 & 2 & 3 & 4\\
\hline 
  Benchmark $P_0$ [dBm] & -37.2 & -29.7 & -25.4 & -28.7 \\
  Benchmark PLE & 3.80 & 3.54 & 3.68 & 3.66  \\
  Benchmark $\alpha$ & 0.54 & 0.54 & 0.56 & 0.53 \\
 \hline
  \# Iterations (initialization) & 1770 & 2571 & 2354 & 2572 \\
  Computation time [ms] & 578  & 459  & 413  & 418 \\
  Initial $P_0$ [dBm] & -45.9 & -35.3 & -29.5 & -30.0  \\
  Initial PLE & 2.86 & 2.86 & 3.19 & 3.34  \\
  Initial $\alpha$ & 0.51 & 0.53 & 0.53 & 0.52 \\\cline{1-5}
  \# Iterations (ranging module) & 98 & 122 & 106 & 97\\
  Computation time [ms] & 818 & 465 & 400 & 437\\
  Last $P_0$ [dBm] & -36.0 & -26.2 & -23.5 & -25.1  \\
  Last PLE & 3.70 & 3.70 & 3.67 & 3.78  \\
  Last $\alpha$ & 0.55 & 0.54 & 0.55 & 0.52 \\
\hline
\end{tabular}
\end{table}

\begin{figure*}
    \captionsetup{farskip=0pt}%
    \centering
    \subfloat[]{\includegraphics[width=0.3\textwidth]{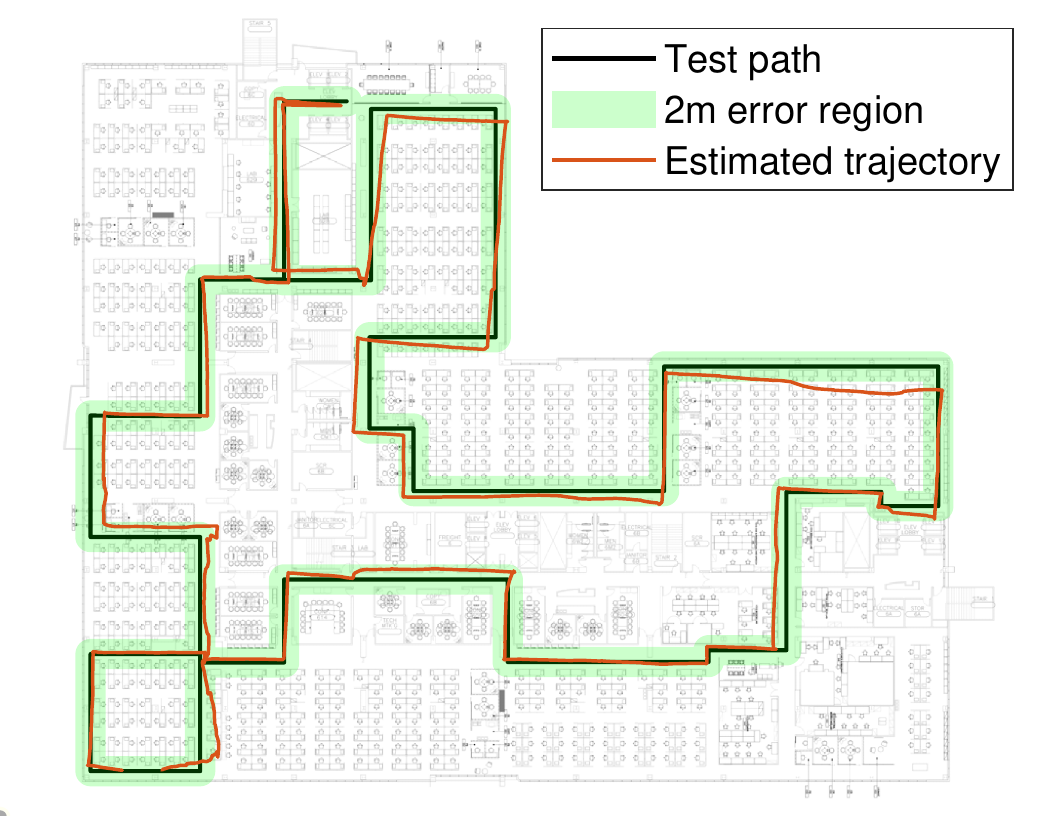}}\hfil\hfil\hfil
    \subfloat[]{\includegraphics[width=0.3\textwidth]{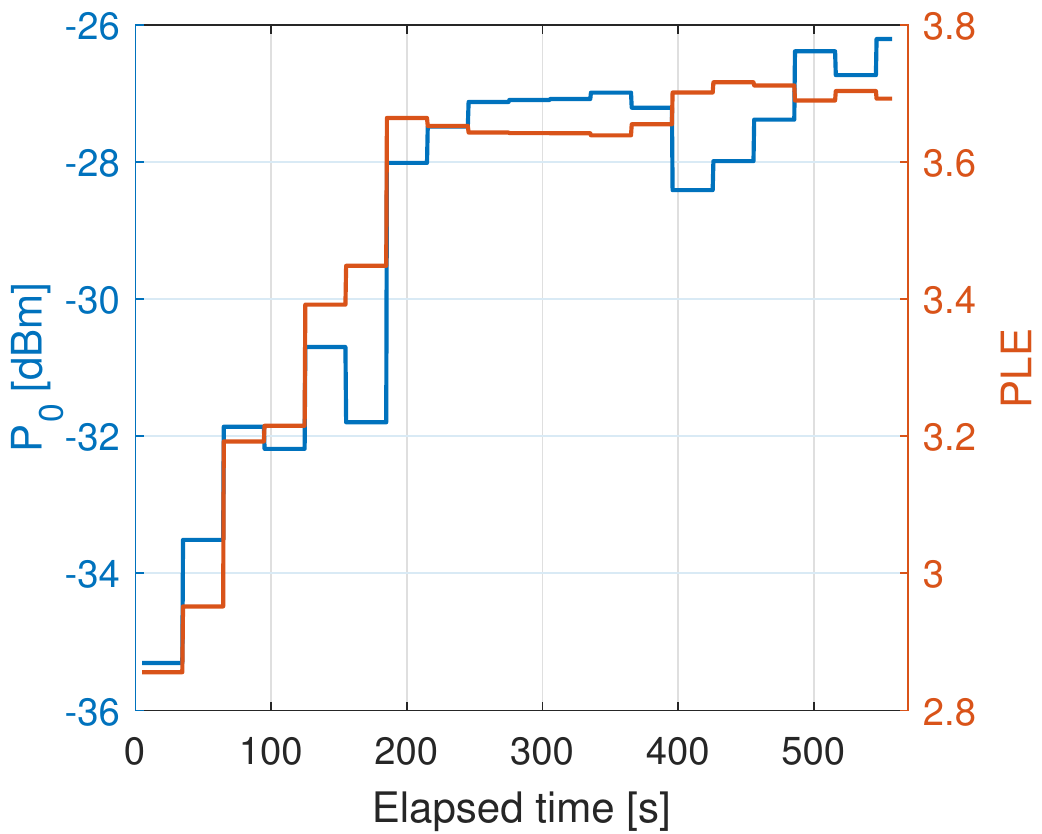}}\hfil\hfil\hfil
    \subfloat[]{\includegraphics[width=0.3\textwidth]{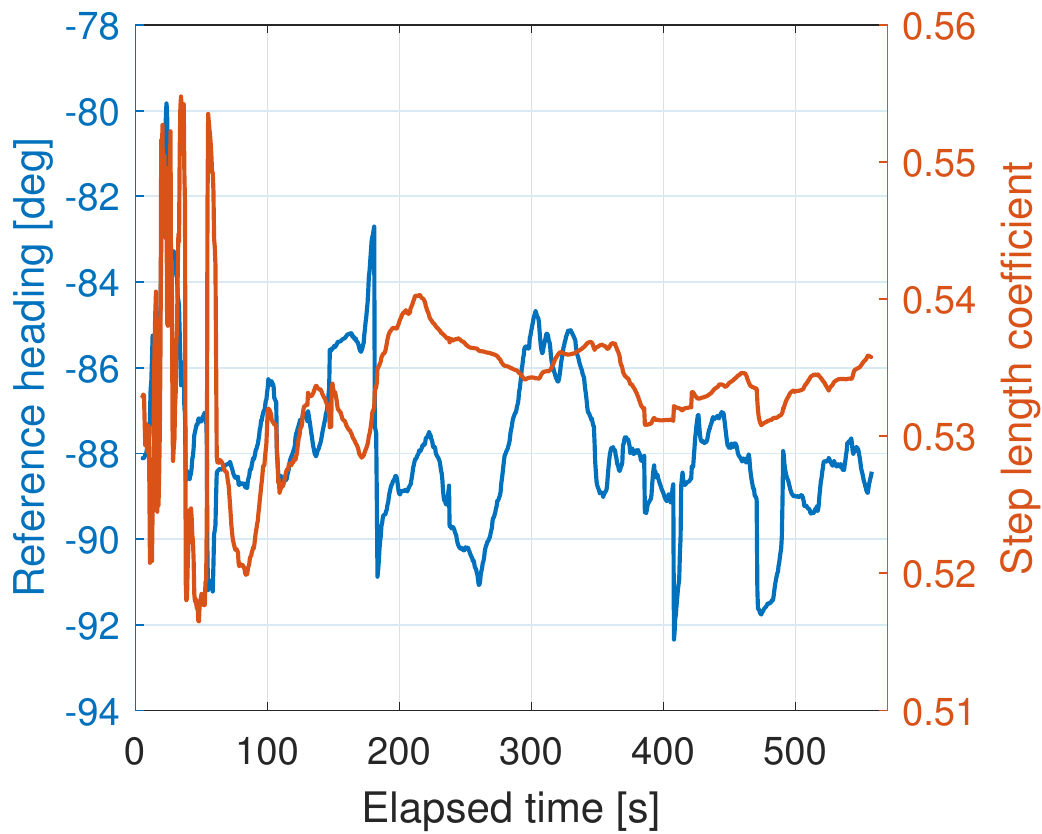}}
    \caption{Example of the proposed online calibration method: (a) estimated trajectory, (b) ranging parameters, and (c) PDR parameters.}
    \label{fig_rss_run}
\end{figure*}

\begin{figure*}
    \captionsetup{farskip=0pt}%
    \centering
    \includegraphics[width=0.3\textwidth]{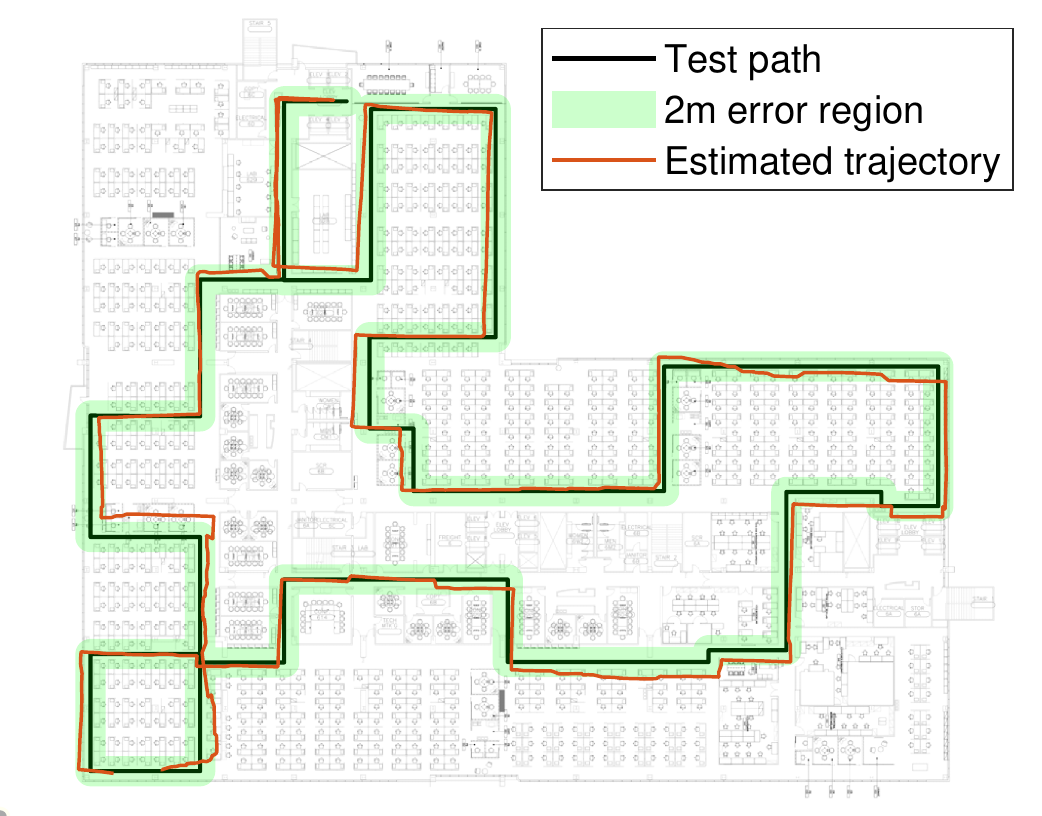}\hfil\hfil\hfil
    \includegraphics[width=0.3\textwidth]{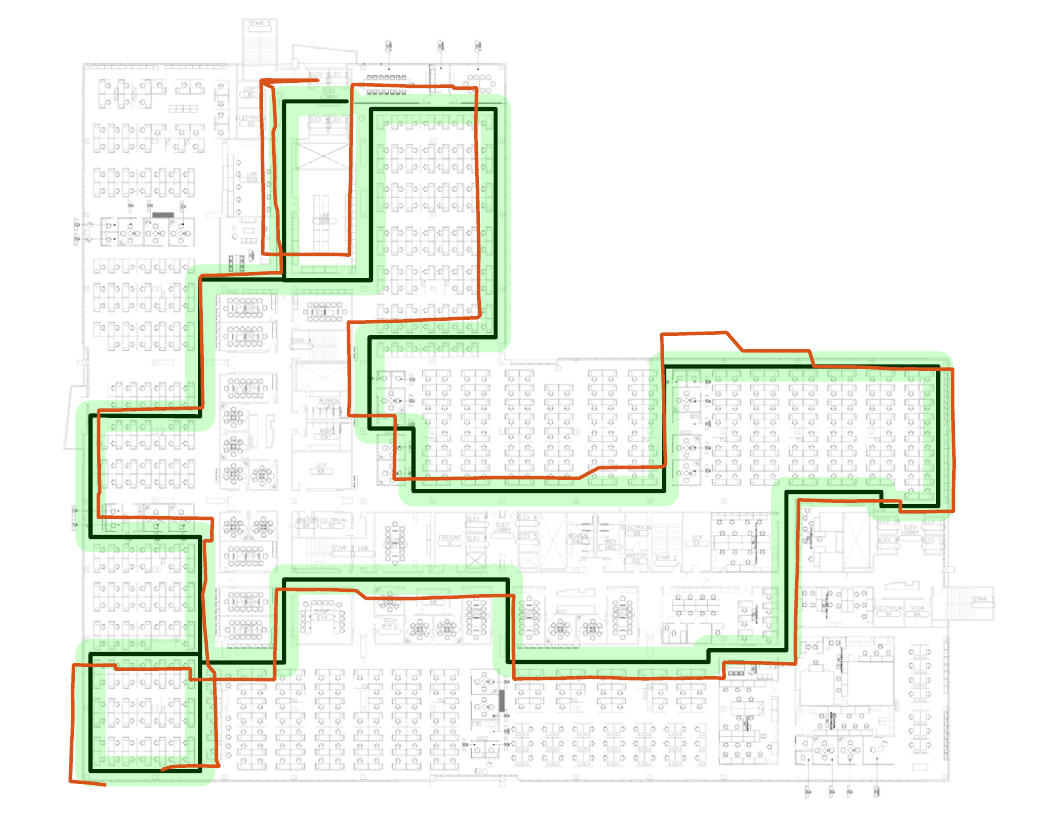}\hfil\hfil\hfil
    \includegraphics[width=0.3\textwidth]{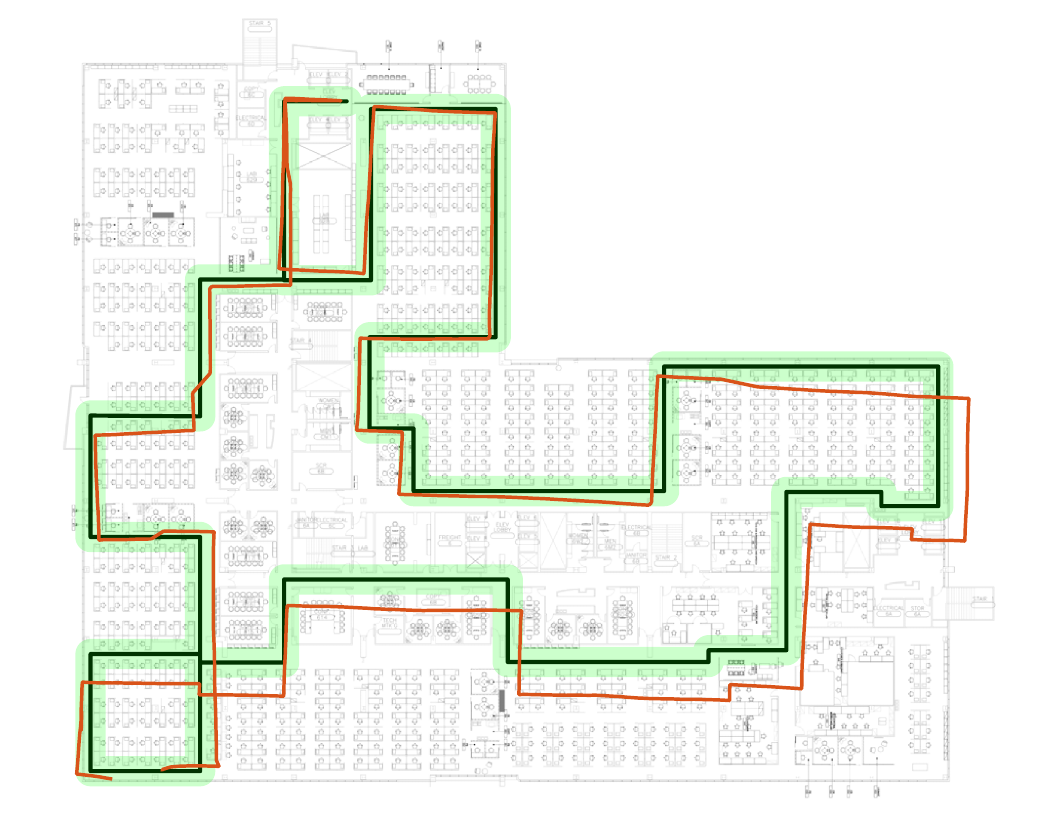}\vfil
    \subfloat[]{\includegraphics[width=0.3\textwidth]{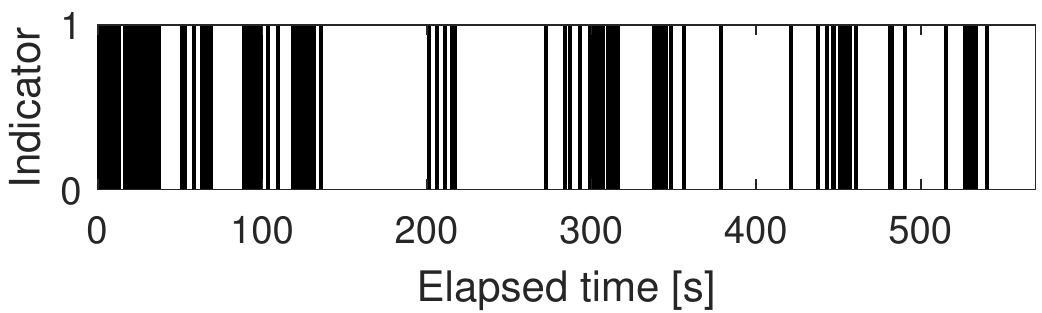}}\hfil\hfil\hfil
    \subfloat[]{\includegraphics[width=0.3\textwidth]{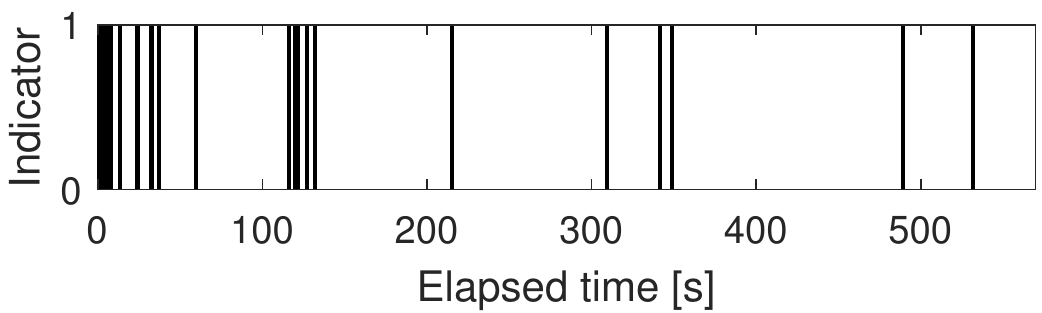}}\hfil\hfil\hfil
    \subfloat[]{\includegraphics[width=0.3\textwidth]{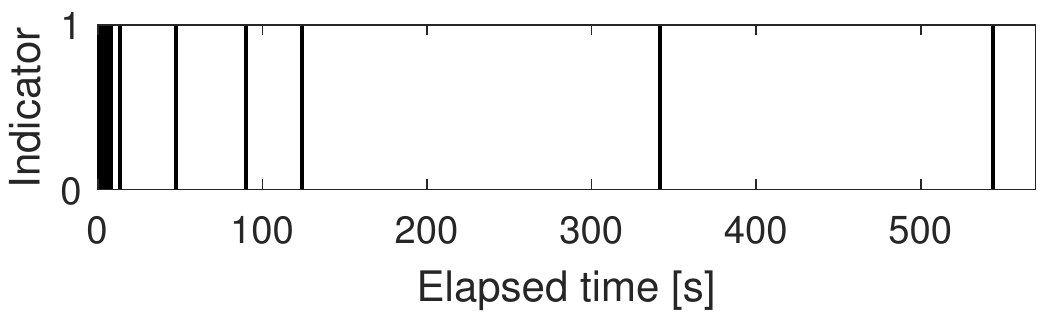}}
    \caption{Positioning performance of the irregular Wi-Fi ranging scenario with different values of threshold: (a) $\rho = 0.2$, (b) $\rho=0.4$, and (c) $\rho=0.8$.}
    \label{fig_rss_run_lambda}
\end{figure*}

To select optimal parameters in the PDR module, we assumed perfect alignment of the reference heading direction and perfect start position of the device.
After that, we selected a step length coefficient $\alpha$ between 0.3 and 0.7 and evaluated the proposed KF-based positioning performance with each selected $\alpha$.
Among the many values of $\alpha$, we selected an optimal one that achieved the best positioning performance in terms of average accuracy.
Fig.~{\ref{fig_rss_benchmark}} illustrates an example of the estimated trajectory of the benchmark scenario for a selected device (i.e., Samsung Galaxy Note 8).
The figure indicates that the start position and the reference heading direction are perfectly given so that the estimated trajectory completely overlaps the test path at the very early stage.
The green area represents the 2~m error region, meaning that all points in this region are up to 2~m away from the nearest point in the test path.
Table~\ref{table_rss_parm} summarizes the choice of optimal parameters for the benchmark scenario.

Unlike the benchmark scenario, the proposed method optimizes every parameter in real-time.
For the initial calibration, the application performs the Wi-Fi ranging procedure when the first $B=8$ steps are detected (This usually takes 4~s).
After the last Wi-Fi ranging procedure, the application conducts the initial calibration procedure described in Section~\ref{sec_init_cal}.
In this experiment, we set the learning rate as $\lambda = 0.001$ and the maximum iterations as 5000.
In addition, if the parameter update can not improve the cost by more than 0.1\% over 10 iterations, the initial calibration is terminated early.

Fig.~\ref{fig_rss_init} illustrates the initial calibration results for the same device used to generate the results shown in Fig.~\ref{fig_rss_benchmark}.
The initial value for each parameter was arbitrary given as $P_0 = -40$~dBm, $\eta = 2$, $\phi_{ref}=0$, and $\alpha=0.5$.
Furthermore, the start position of the device was simply initialized as the coordinates of the AP with the strongest RSS measurement.
According to the early stopping criteria, the initial calibration procedure was stopped at 2571 iterations, which took only 427~ms on the device. 
Fig.~\ref{fig_rss_init}(a) shows the estimated trajectory after 0, 50, 100, 800, and 2500 iterations.
The estimated trajectory of the device gradually approaches to the test path as the number of iterations increases.
Fig.~\ref{fig_rss_init}(b) indicates that the cost decreases with the iteration as parameters in the ranging module and the PDR module are optimized, as shown in Fig.~\ref{fig_rss_init}(c) and (d), respectively.
After the initial calibration, the parameters for the device were given by $P_0=-35.3$ dBm, $\eta = 2.86$, and $\alpha = 0.53$.
The initial calibration results for other devices are also summarized in Table~\ref{table_rss_parm}.

\begin{figure*}
    \centering
    \captionsetup{farskip=0pt}%
    \subfloat[]{\includegraphics[width=0.3\textwidth]{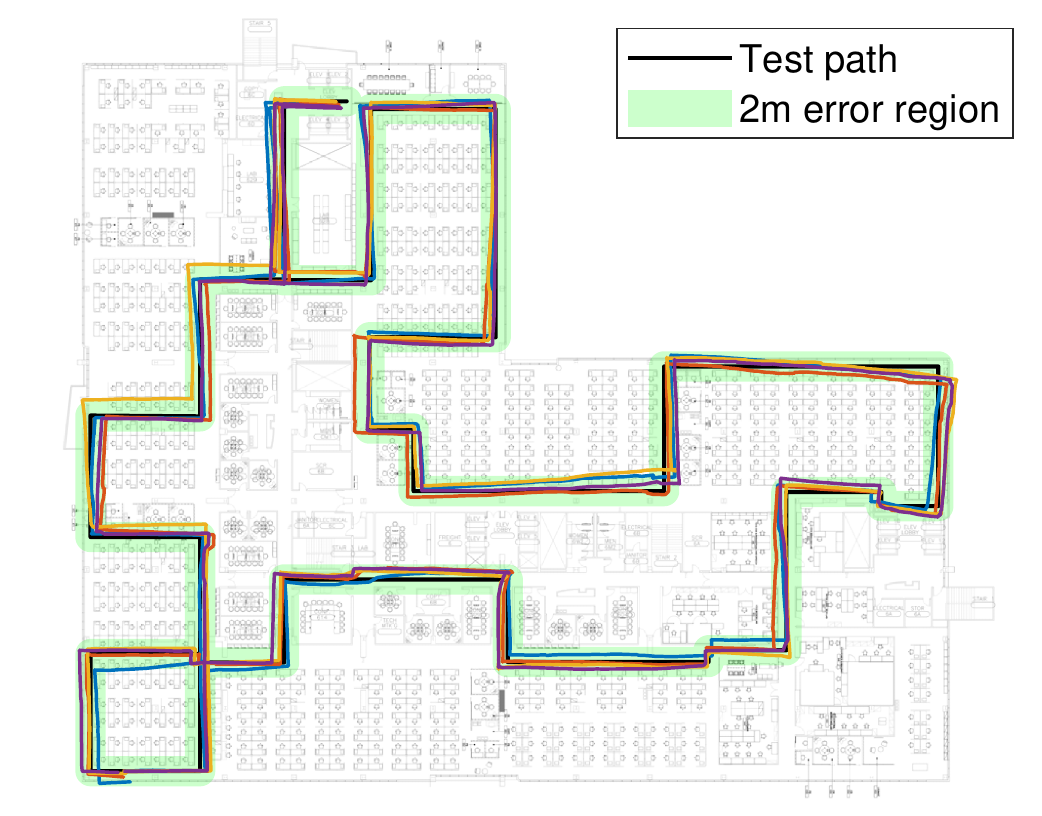}}\hfil\hfil\hfil
    \subfloat[]{\includegraphics[width=0.3\textwidth]{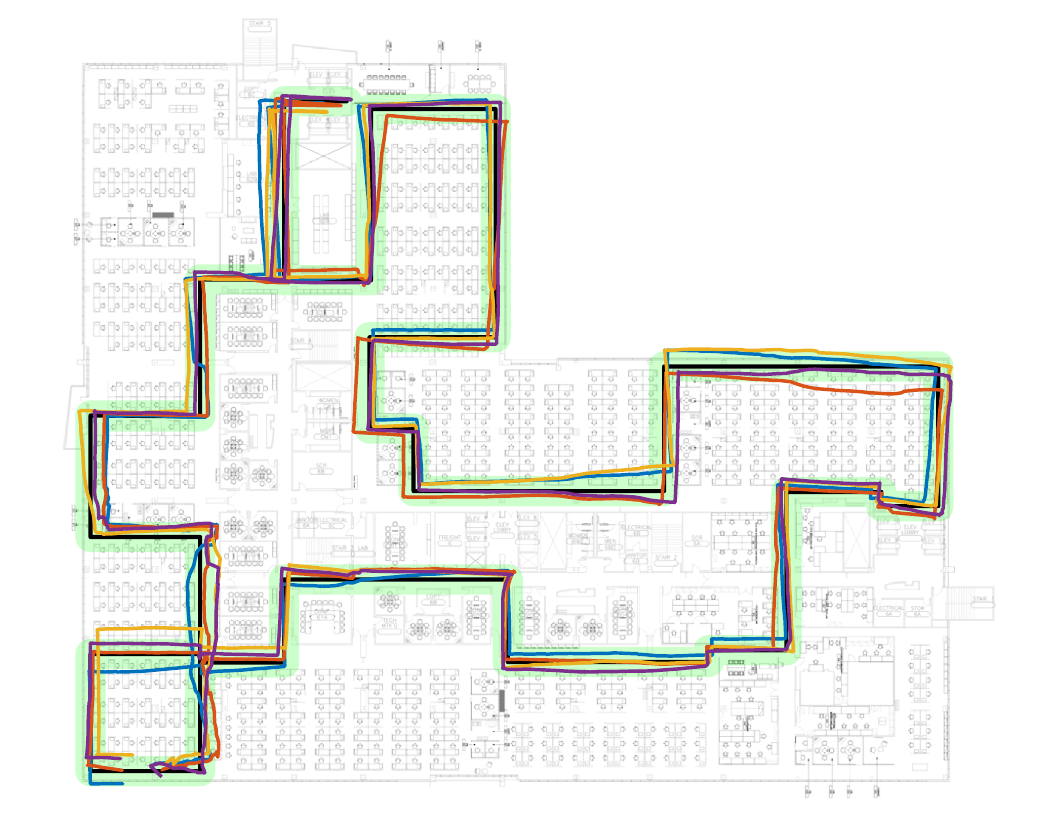}}\hfil\hfil\hfil
    \subfloat[]{\includegraphics[width=0.3\textwidth]{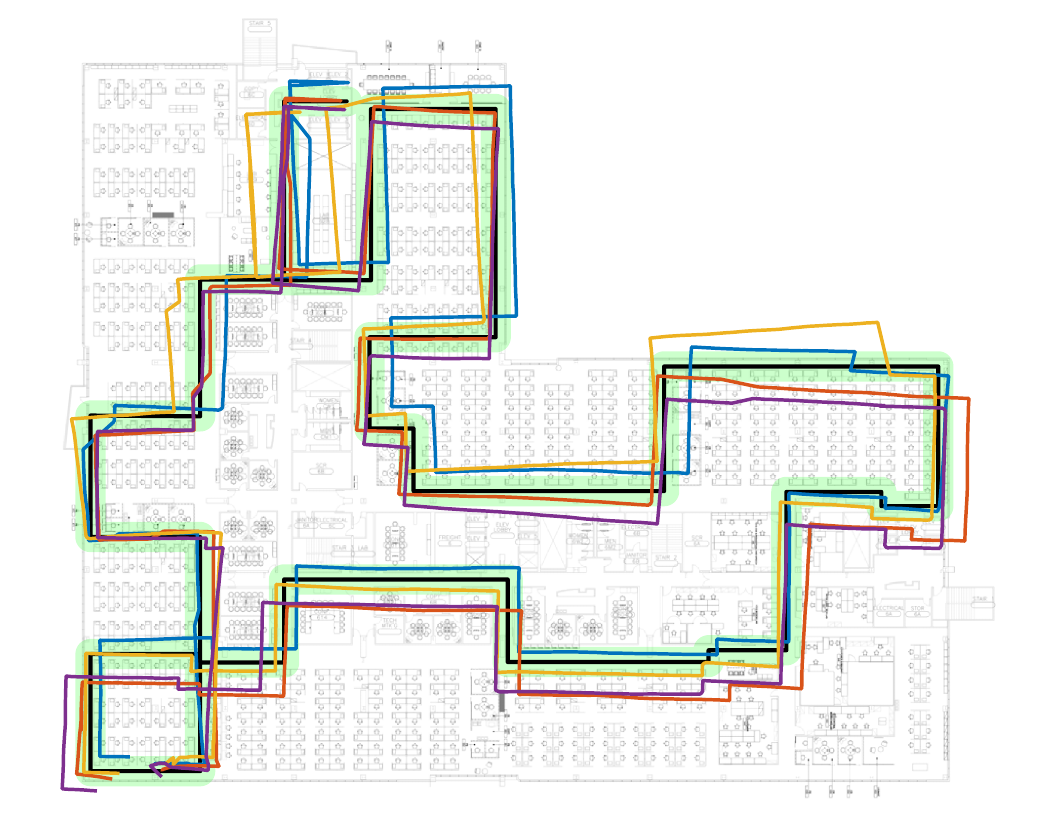}}
    \caption{Estimated trajectories of all devices: (a) benchmark scenario, (b) proposed method with $\rho=0$, and (c) proposed method with $\rho=0.8$.}
    \label{fig_rss_all}
\end{figure*}

\begin{figure*}
    \centering
    \captionsetup{farskip=0pt}%
    \subfloat[]{\includegraphics[width=0.3\textwidth]{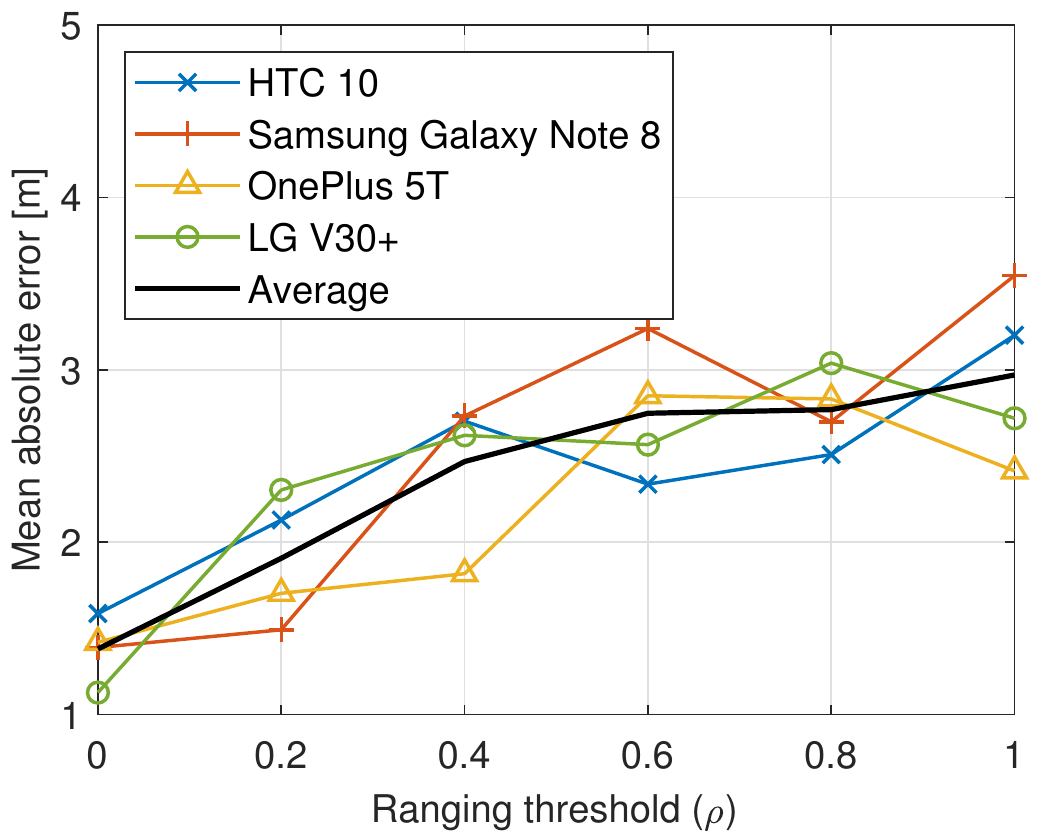}}\hfil\hfil\hfil
    \subfloat[]{\includegraphics[width=0.3\textwidth]{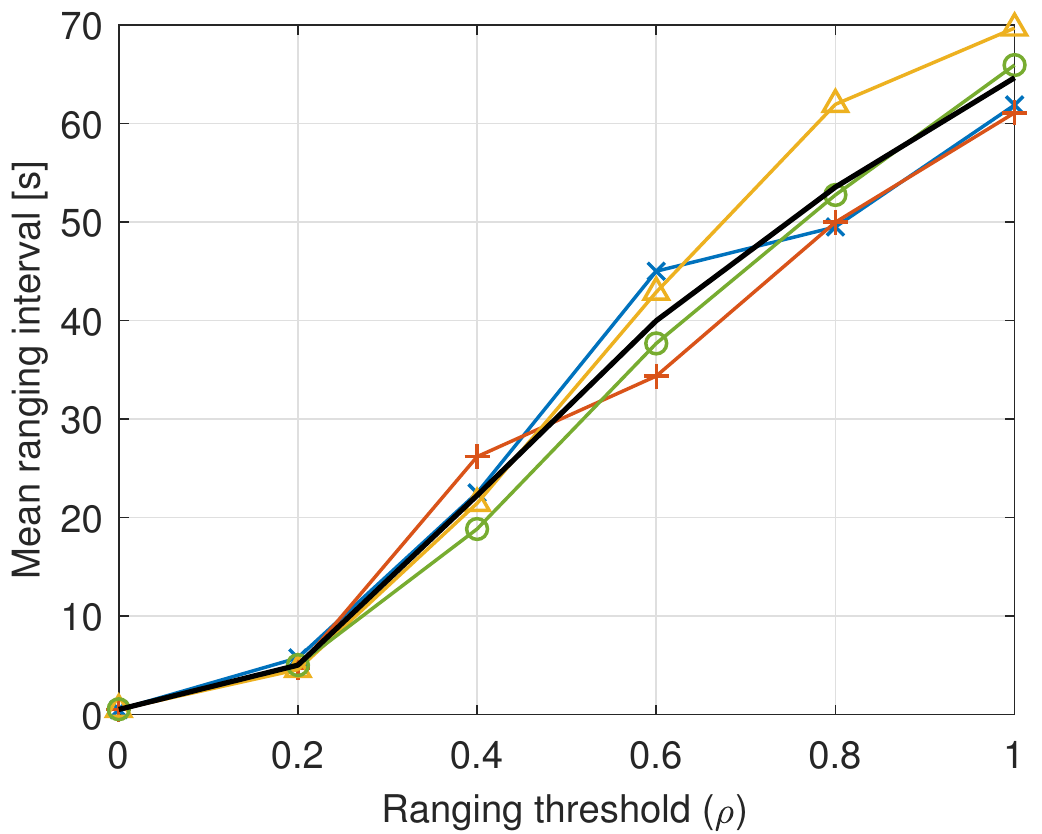}}\hfil\hfil\hfil
    \subfloat[]{\includegraphics[width=0.3\textwidth]{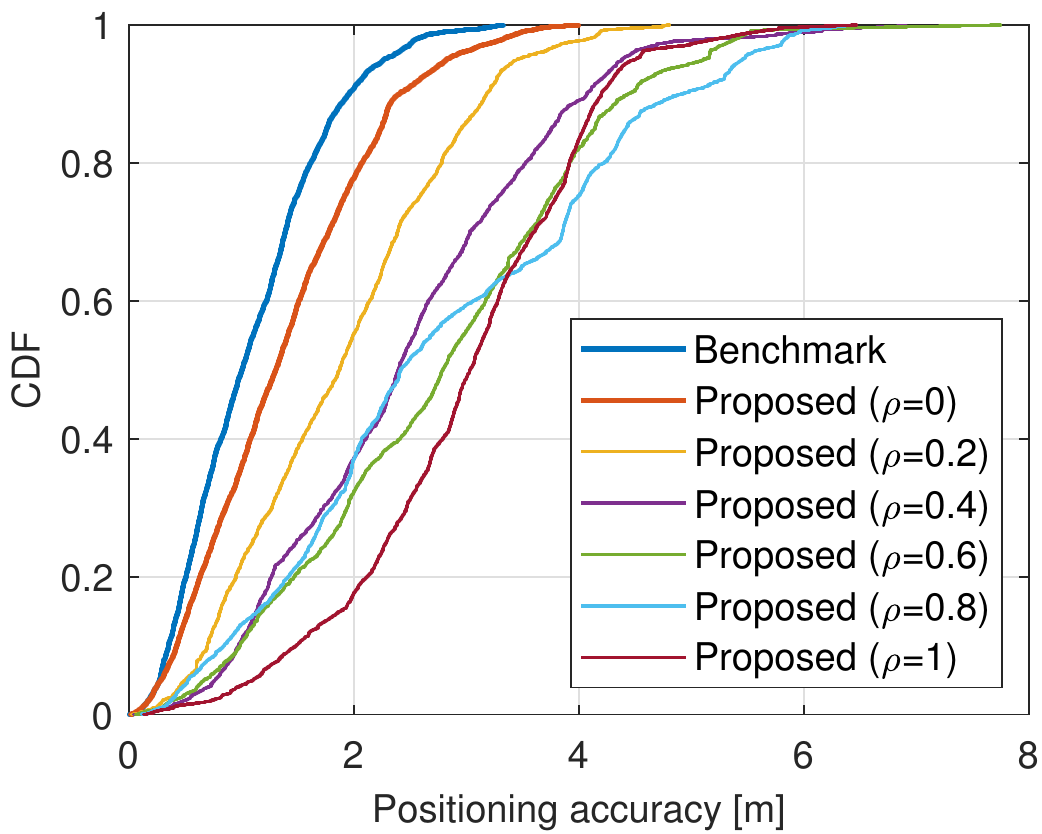}}
    \caption{Performance depending on the ranging threshold: (a) mean absolute error, (b) mean ranging interval, and (c) cumulative density function of positioning accuracy.}
    \label{fig_rss_cdf}
\end{figure*}

Beginning with the initial parameter estimates, the application continues to update all parameters.
Using the same assumption as the benchmark scenario, we first considered an excessive Wi-Fi ranging scenario by simply assigning $\rho=0$.
The application executes the self-calibration procedure for the ranging module every 30~s; thus, the ranging parameters were updated relatively slowly, as shown in Fig.~\ref{fig_rss_run}(b).
However, the parameters in the PDR module are updated whenever new Wi-Fi ranging results are available.
Therefore, these parameters were updated more frequently, as shown in Fig.~\ref{fig_rss_run}(c).

To improve memory efficiency and computation complexity, we used only up to 100 latest Wi-Fi ranging results for the self-calibration procedure.
Furthermore, the same learning rate and the same early stopping criteria were applied for the ranging calibration.
In general, each ranging calibration procedure completed within relatively short iterations as they began with the previously optimized parameters.
For instance, the self-calibration procedures completed in 123 iterations on average, which took 427~ms on average on the device.
The average ranging calibration iterations and computation time for each device are summarized in Table~\ref{table_rss_parm}.
In addition, the lastly updated parameters are also summarized in the table.

In addition to the excessive Wi-Fi ranging scenario, we evaluated the positioning performance using irregular ranging scenarios.
Fig.~\ref{fig_rss_run_lambda}(a), (b), and (c) depict the estimated trajectory of the device with different values of $\rho$. 
The sub-figure located under each estimated trajectory represents the ranging procedure indicators, where the black bar indicates that the ranging procedure was executed.
The larger the threshold used, the less often the ranging procedure was executed, which improves battery life and network performance while sacrificing the positioning accuracy.
Other devices addressed the same trend.
Fig.~\ref{fig_rss_all} depicts the estimated trajectories of all devices used in the experiment for the benchmark scenario and the proposed method with $\rho=0$ and $0.8$ scenarios.


\begin{table}
\caption{Positioning Performance with RSS-based Ranging and PDR}
\label{table_rss_result}
\renewcommand{\arraystretch}{1.2}
\centering
\begin{tabular}{c|c|cccc|c}
 & Device \# & 1 & 2 & 3 & 4 & Mean\\
\hline 
{\multirow{4}{*}{\rotatebox[origin=c]{90}{Benchmark}}} & Mean ranging interval [s] & 0.5  & 0.5 & 0.5  & 0.5  & 0.5\\
 & MAE [m]              & 1.15 & 0.99 & 1.14 & 1.09 &  1.09\\
 & RMSE [m]             & 1.31 & 1.17 & 1.34 & 1.22 &  1.26\\
 & 75\%-tile error [m]  & 1.51 & 1.46 & 1.43 & 1.52 &  1.48\\
 \hline
{\multirow{4}{*}{\rotatebox[origin=c]{90}{Proposed}}}{\multirow{4}{*}{\rotatebox[origin=c]{90}{$\rho=0$}}} & Mean ranging interval [s] & 0.5 & 0.5 & 0.5 & 0.5 & 0.5\\
 & MAE [m]              & 1.59 & 1.39 & 1.42 & 1.13 &  1.38\\
 & RMSE [m]             & 1.75 & 1.61 & 1.66 & 1.31 &  1.58\\
 & 75\%-tile error [m]  & 2.14 & 1.90 & 2.09 & 1.52 &  1.91\\
 \hline
{\multirow{4}{*}{\rotatebox[origin=c]{90}{Proposed}}}{\multirow{4}{*}{\rotatebox[origin=c]{90}{$\rho=0.2$}}} & Mean ranging interval [s] & 5.8 & 4.7 & 4.6 & 5.0 & 5.0\\
 & MAE [m]              & 2.13 & 1.50 & 1.70 & 2.30 &  1.91\\
 & RMSE [m]             & 2.35 & 1.65 & 1.92 & 2.51 &  2.11\\
 & 75\%-tile error [m]  & 2.74 & 1.93 & 2.40 & 3.03 &  2.53\\
 \hline
{\multirow{4}{*}{\rotatebox[origin=c]{90}{Proposed}}}{\multirow{4}{*}{\rotatebox[origin=c]{90}{$\rho=0.4$}}} & Mean ranging interval [s] & 22.5 & 26.2 & 21.4 & 18.8  &  22.2\\
 & MAE [m] & 2.70 & 2.73 & 1.82 & 2.62  & 2.47\\
 & RMSE [m] & 3.09 & 2.92 & 2.04 & 2.83   & 2.72\\
 & 75\%-tile error [m] & 3.66 & 3.47 & 2.45 & 3.49   & 3.27 \\
 \hline
{\multirow{4}{*}{\rotatebox[origin=c]{90}{Proposed}}}{\multirow{4}{*}{\rotatebox[origin=c]{90}{$\rho=0.8$}}} & Mean ranging interval [s] & 49.5 & 50.0 & 61.9 & 52.7  &  53.5 \\
 & MAE [m] &  2.51 & 2.70 & 2.83 & 3.04   &  2.77\\
 & RMSE [m] & 2.77 & 3.26 & 3.18 & 3.56  &  3.14\\
 & 75\%-tile error [m] & 3.50 & 4.38 & 3.95 & 4.34 & 4.04\\
 \hline
\end{tabular}
\end{table}

We can quantitatively analyze the performance depending on the value of the ranging threshold.
The main positioning accuracy evaluation metric in this paper is the mean absolute error (MAE), which is defined by $MAE = E[\lVert\hat{\mathbf p} - \mathbf p^*\rVert]$ with the true position of the device $\mathbf p^*$.
In addition, the root mean squared error (RMSE), which is defined by $RMSE = \sqrt{E[\lVert\hat{\mathbf p} - \mathbf p^*\rVert^2]}$, and the 75 percentile accuracy can also be used to evaluate the positioning accuracy.
The other evaluation metric can be the mean ranging interval.
Note that the burst Wi-Fi ranging procedures for the initial calibration are not counted in the computation of the mean ranging interval.

Fig.~\ref{fig_rss_cdf}(a) shows that the MAE of each device tends to increase with the ranging threshold. 
This is because a larger $\rho$ invokes Wi-Fi ranging procedure less frequently, as shown in Fig.~\ref{fig_rss_cdf}(b).
Between consecutive Wi-Fi ranging procedures, the positioning of the device relies solely on the PDR module based on the lastly updated parameters.
Hence, if the latest estimate of the reference heading direction is not accurate, the PDR module generates a slightly wrong trajectory until it is corrected with the next Wi-Fi ranging results.
Fig.~\ref{fig_rss_cdf}(c) shows the cumulative density function (CDF) of positioning accuracy. For this figure, positioning results from all devices were combined together for each CDF curve.
For the excessive Wi-Fi ranging scenario, the proposed method closely achieved the benchmark positioning result, even without calibrating any parameters in advance.
As a larger $\rho$ was applied, the CDF curve shifts right from the benchmark CDF.
Finally, the positioning performance of RSS-based ranging and PDR is summarized in Table~\ref{table_rss_result}.

\subsection{Positioning with RTT-Based Ranging and PDR}

Subsequently, we verified the performance of RTT-based positioning with newly installed FTMRs and the Google Pixel devices. 
As a default setting, these devices exchange FTM packets eight times for a single ranging request and report the average ranging results as a measured distance between the devices.
Similar to the previous experiments, we first evaluated the performance of a benchmark scenario, where every ranging and PDR-related parameter is perfectly selected.
To calibrate the FTM protocol, we used a linear calibration polynomial for simplicity. Therefore, the set of ranging parameters is given by $\Theta_{RTT}=\{c_1, c_0\}$. For the benchmark scenario, we selected optimal coefficients to have the minimum MSE between the calibrated and actual distances.
In addition, the reference heading direction and step length coefficient were selected in the same manner from the previous experiment.
We also assumed the excessive ranging scenario that performs ranging procedure every 500~ms for the benchmark scenario.

\begin{figure}
    \captionsetup{farskip=0pt}%
    \centering
    \subfloat[]{\includegraphics[width=0.225\textwidth]{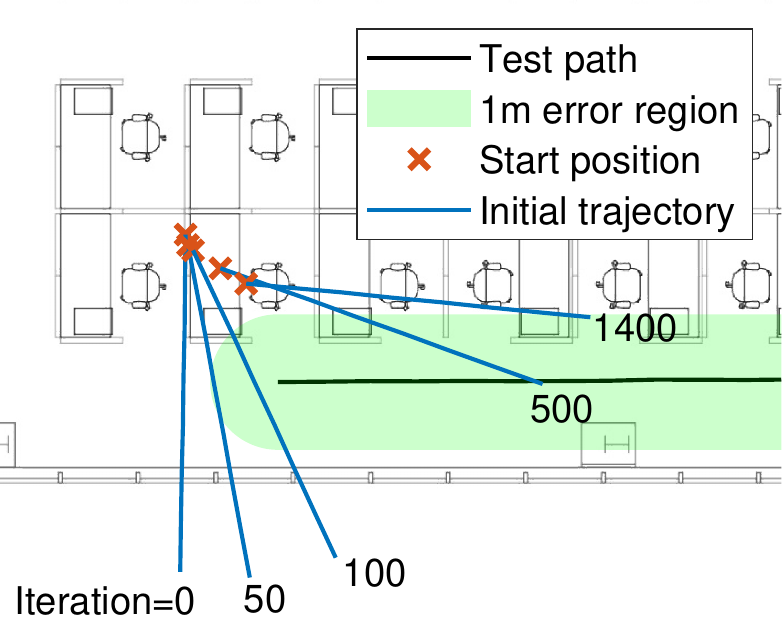}}\hfil
    \subfloat[]{\includegraphics[width=0.225\textwidth]{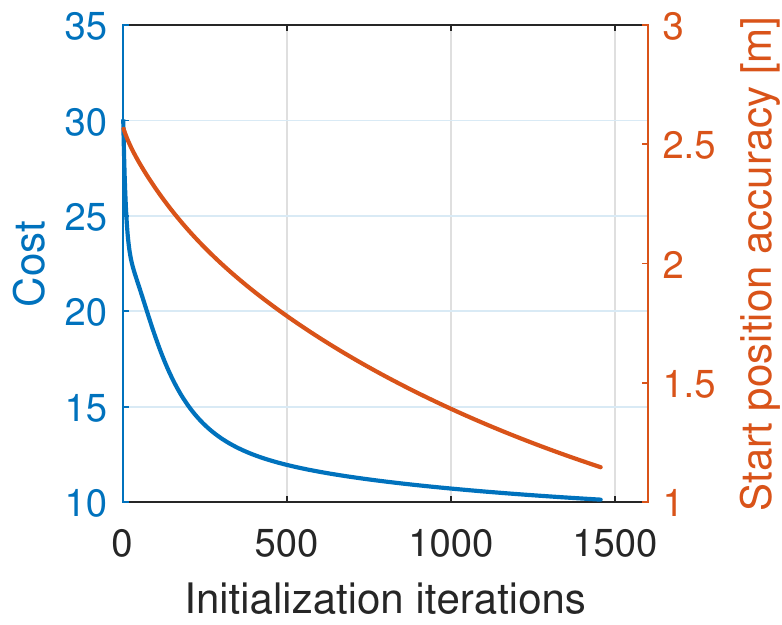}}\vfil
    \subfloat[]{\includegraphics[width=0.225\textwidth]{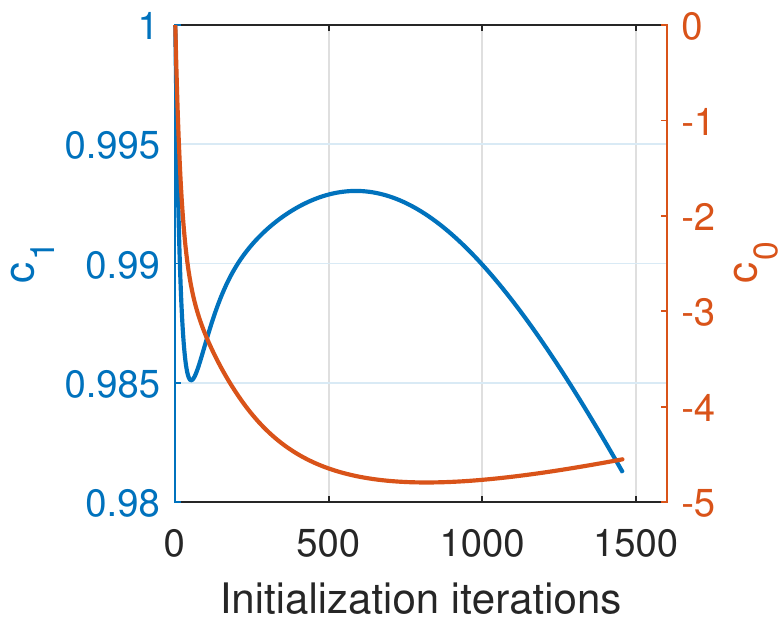}}\hfil
    \subfloat[]{\includegraphics[width=0.225\textwidth]{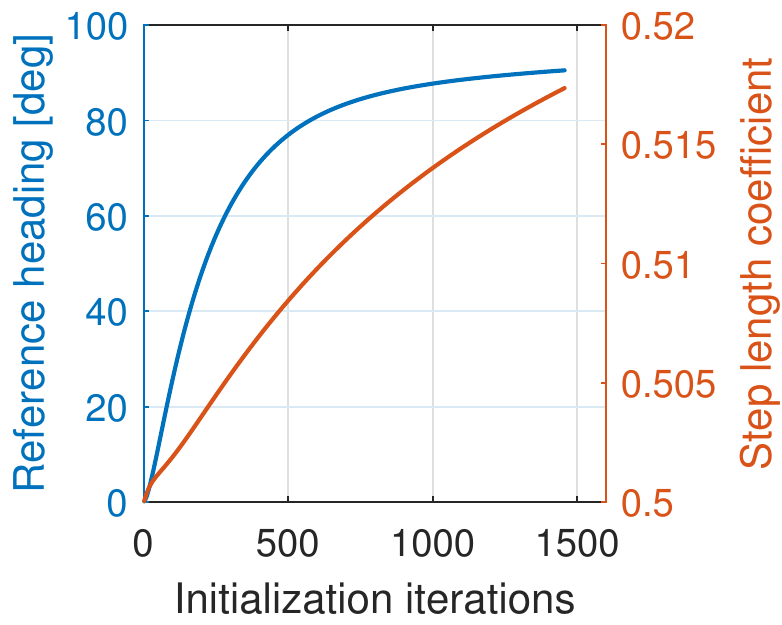}}
    \caption{An example of initial calibration step: (a) estimated trajectory, (b) cost function over time, (c) ranging parameters over time, and (d) PDR parameters over time.}
    \label{fig_rtt_init}
\end{figure}

\begin{table}
\caption{Benchmark Parameters and Online Calibration Results for RTT-based Ranging Scenario}
\label{table_rtt_parm}
\renewcommand{\arraystretch}{1.2}
\centering
\begin{tabular}{c|cccc}
 Device \# & 5 & 6 & 7 & 8\\
\hline 
 Benchmark $c_1$ & 0.81 & 0.85 & 0.87 & 0.94 \\
 Benchmark  $c_0$ & -1.40 & -2.73 & -2.29 & -5.00 \\
 Benchmark  $\alpha$ & 0.53 & 0.55 & 0.53 & 0.54 \\
 \hline
 \# Iterations (initialization) & 2638 & 2745 & 1457 & 2232\\
 Computation time [ms] & 368 & 268 & 98 & 121\\
 Initial $c_1$ & 0.84 & 0.93 & 0.98 & 0.89  \\
 Initial $c_0$ & -1.36 & -4.61 & -4.55 & 2.64  \\
 Initial $\alpha$ & 0.54 & 0.55 & 0.52 & 0.53 \\
 \hline
 \# Iterations (ranging module) & 86 & 181 & 194 & 87\\
 Computation time [ms] & 376  & 671  & 564  & 195 \\
 Last $c_1$ & 0.82 & 0.85 & 0.87 & 0.85  \\
 Last $c_0$ & -3.42 & -3.86 & -3.36 & -4.03  \\
 Last $\alpha$ & 0.51 & 0.55 & 0.54 & 0.52 \\
\hline
\end{tabular}
\end{table}

\begin{figure}
\centering
\subfloat[]{\includegraphics[width=0.42\textwidth]{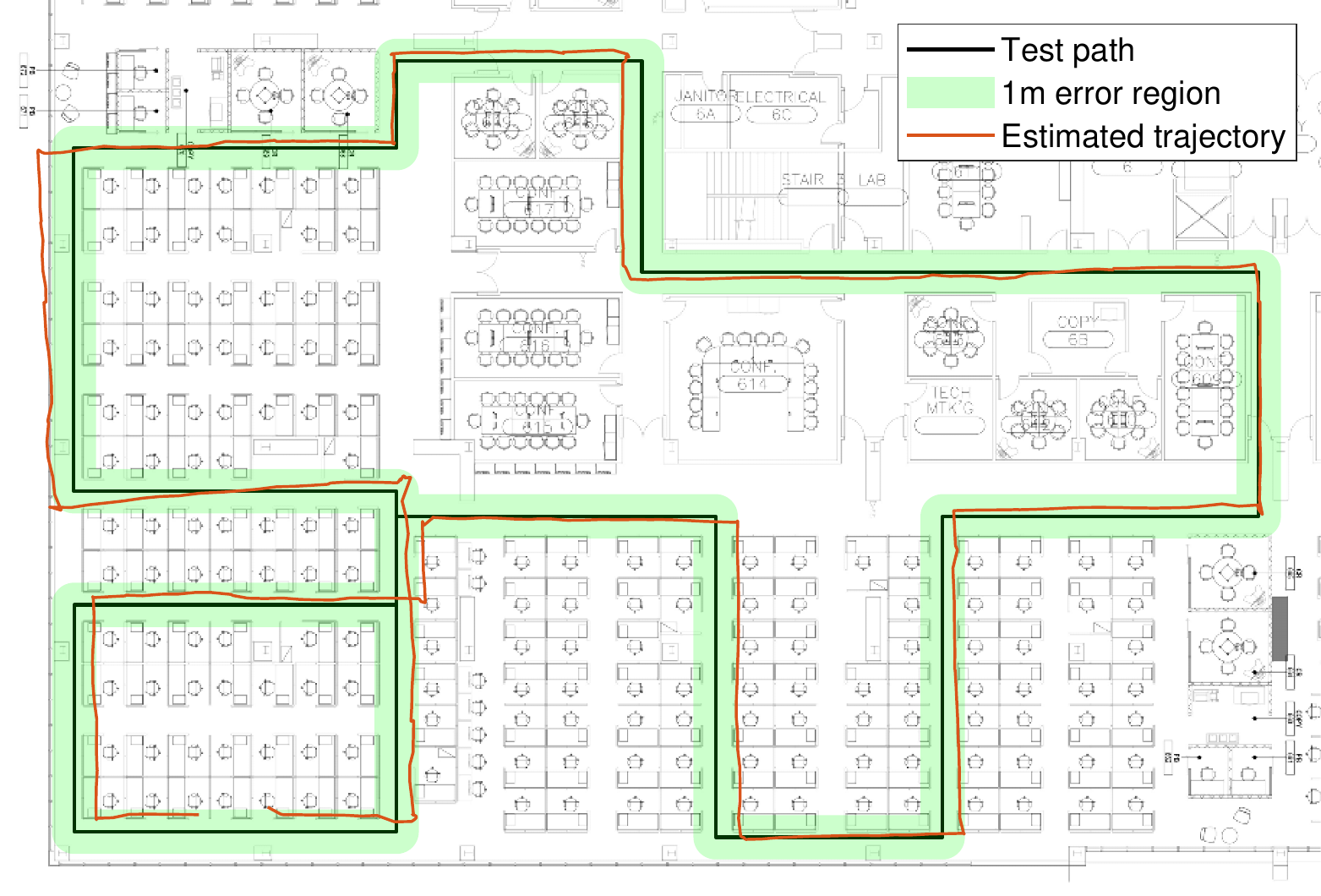}}\\
\subfloat[]{\includegraphics[width=0.21\textwidth]{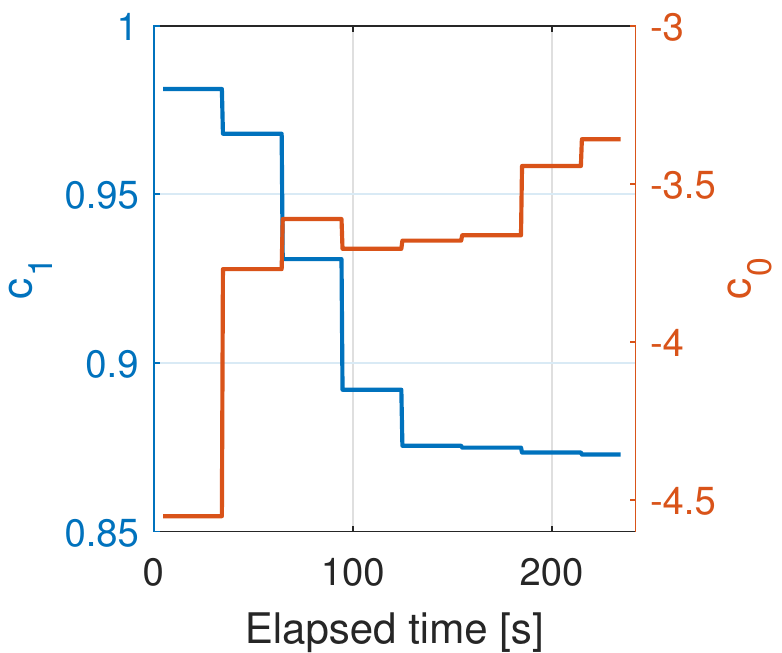}}\hfil\hfil\hfil
\subfloat[]{\includegraphics[width=0.21\textwidth]{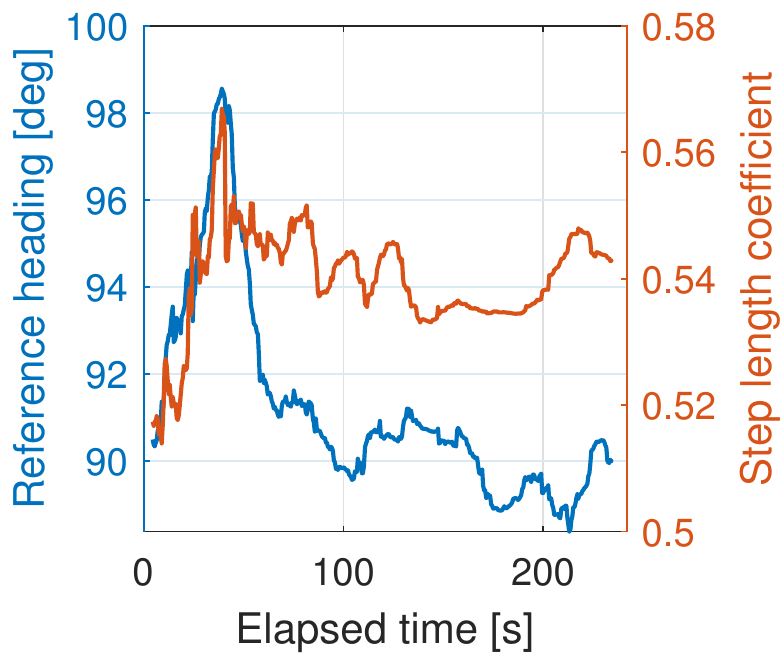}}
\caption{An example of initial calibration step: (a) estimated trajectory, (b) ranging parameters over time, and (c) PDR parameters over time.}
\label{fig_rtt_run}
\end{figure}

\begin{figure*}
    \captionsetup{farskip=0pt}%
    \centering
    \subfloat[]{\includegraphics[width=0.3\textwidth]{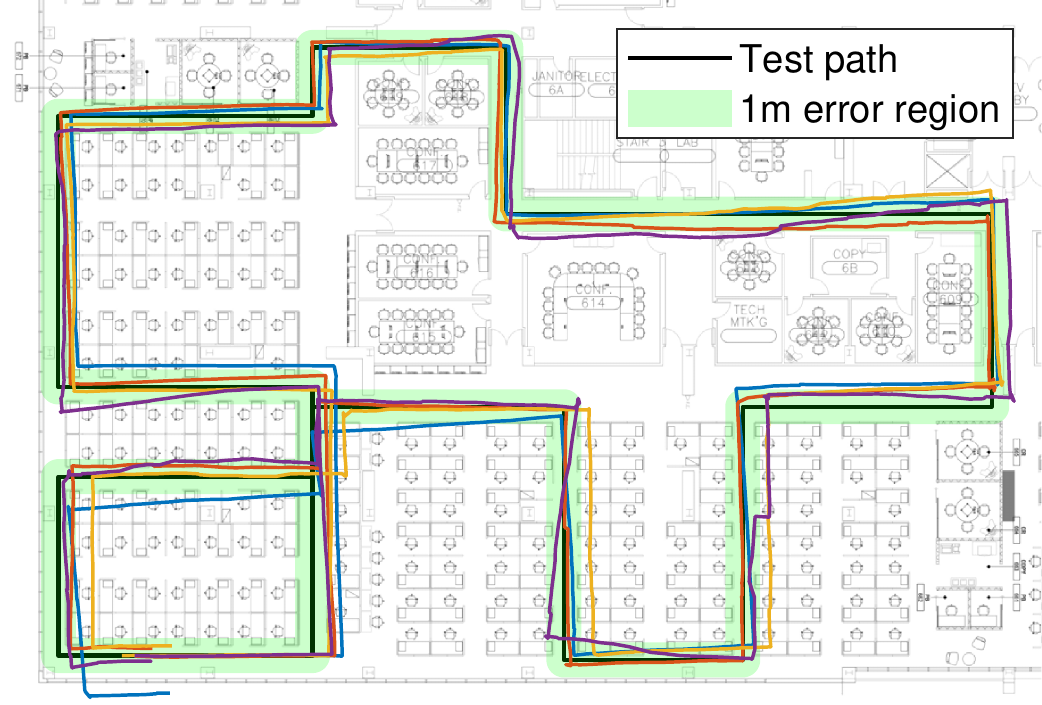}}\hfil\hfil\hfil
    \subfloat[]{\includegraphics[width=0.3\textwidth]{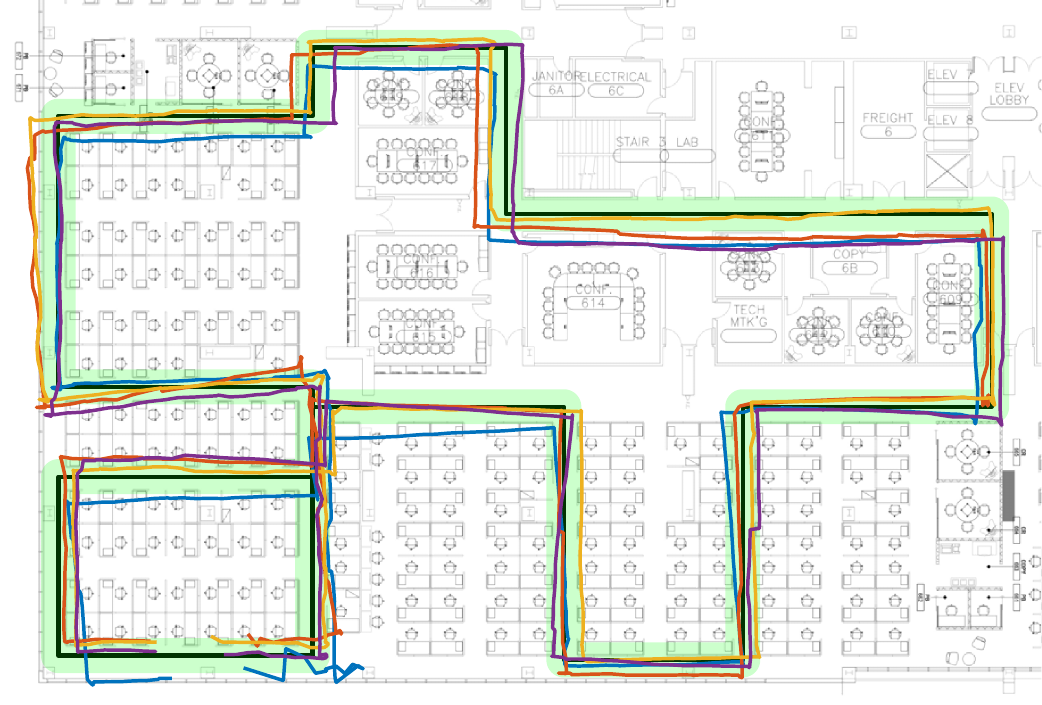}}\hfil\hfil\hfil
    \subfloat[]{\includegraphics[width=0.3\textwidth]{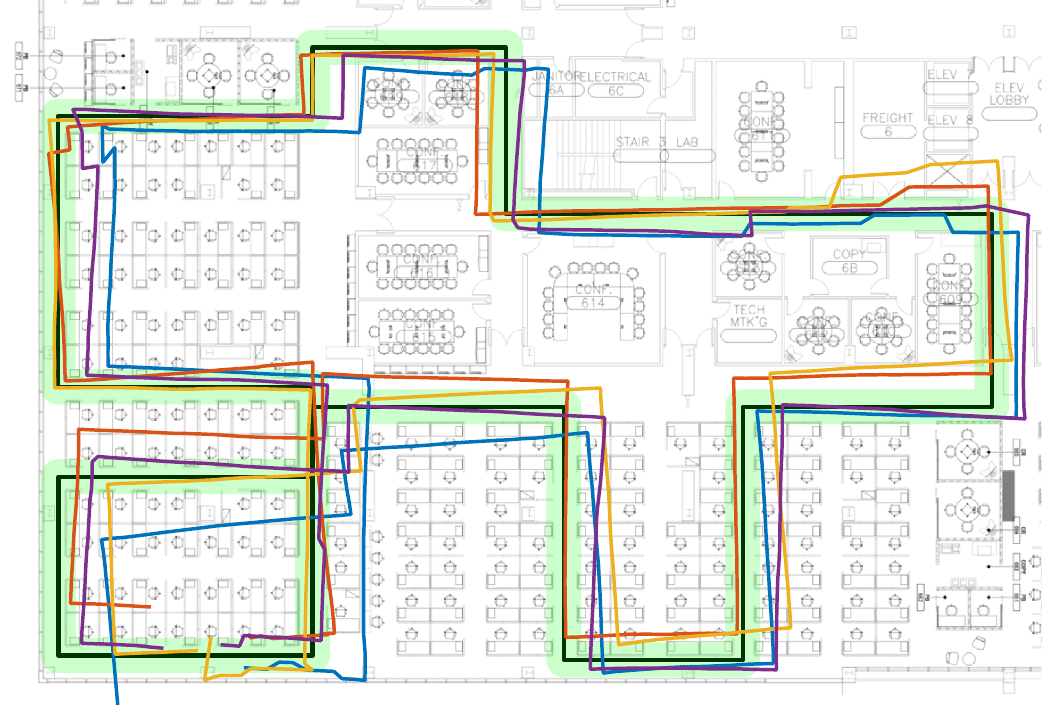}}
    \caption{Estimated trajectories for all devices: (a) benchmark scenario, (b) online calibration scenario with $\rho=0$, and (c) $\rho=0.8$.}
    \label{fig_rtt_all}
\end{figure*}

\begin{figure*}
    \captionsetup{farskip=0pt}%
    \centering
    \subfloat[]{\includegraphics[width=0.3\textwidth]{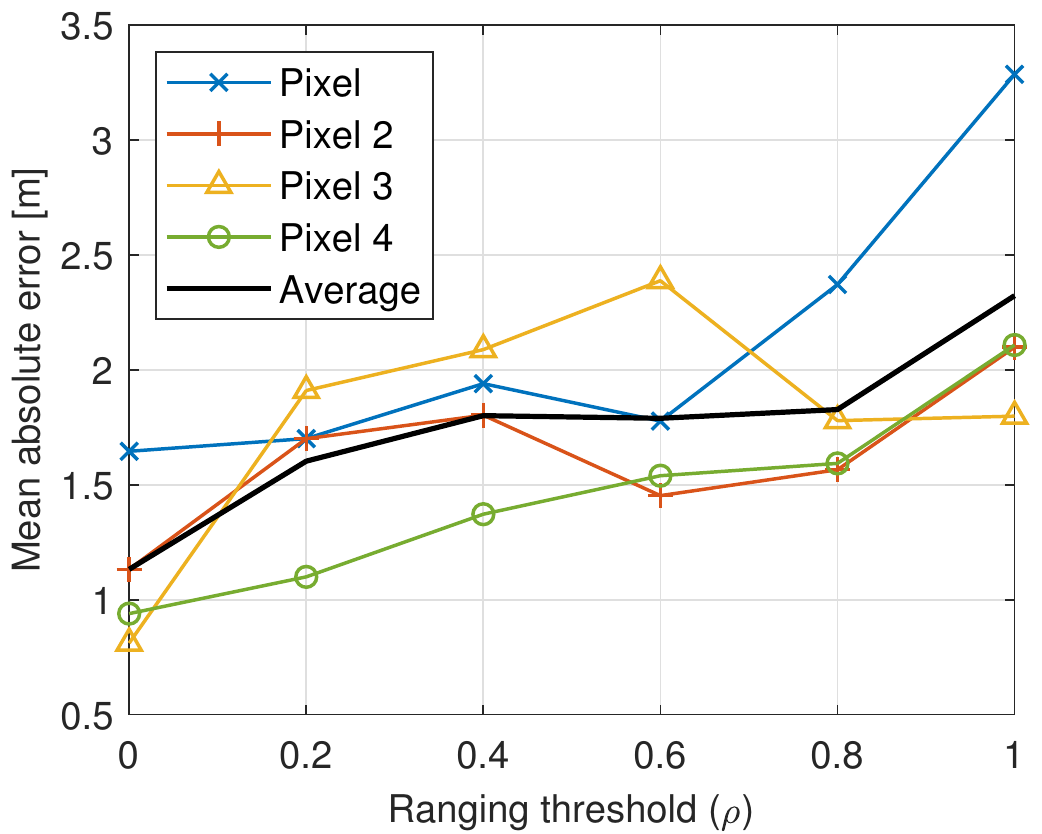}}\hfil\hfil\hfil
    \subfloat[]{\includegraphics[width=0.3\textwidth]{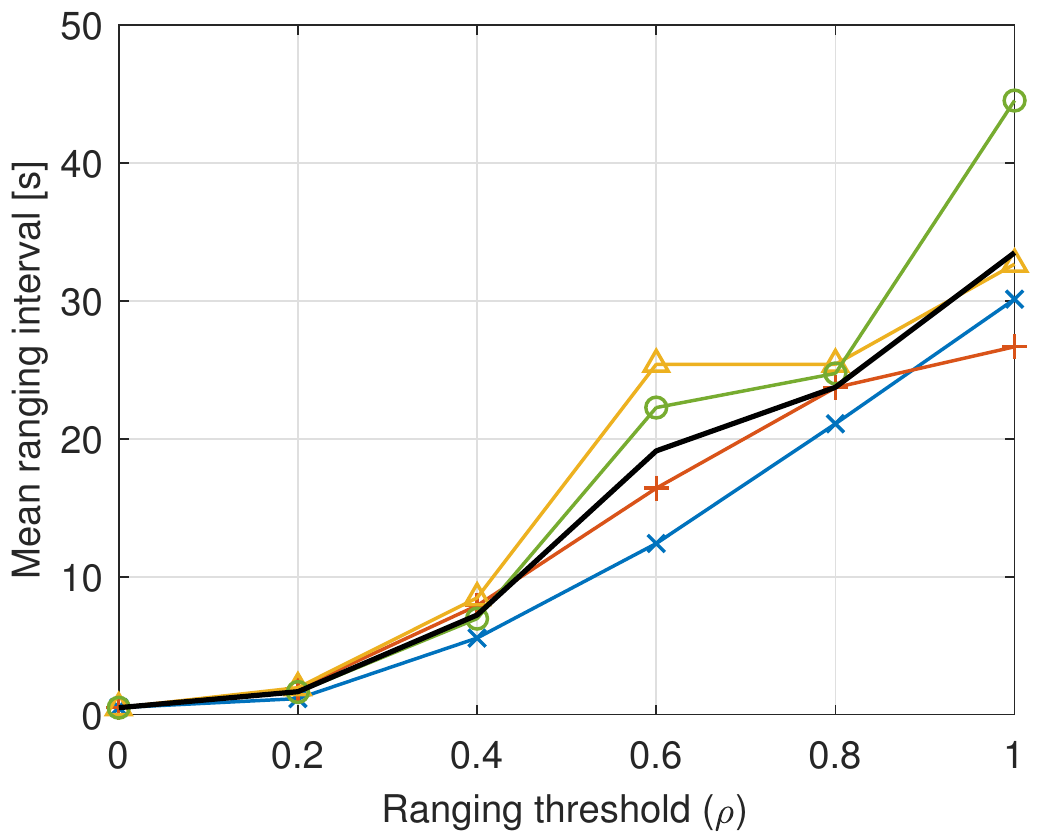}}\hfil\hfil\hfil
    \subfloat[]{\includegraphics[width=0.3\textwidth]{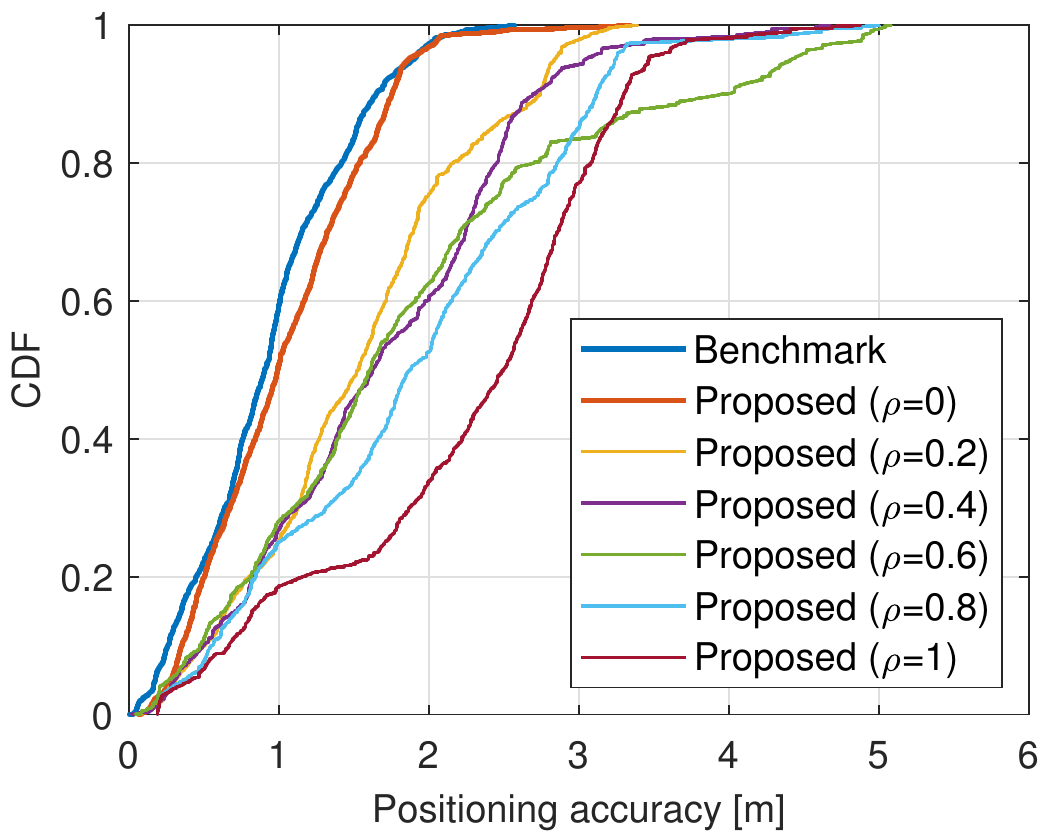}}
    \caption{Performance depending on the ranging threshold: (a) mean absolute error, (b) mean ranging interval, and (c) cumulative density function of positioning accuracy.}
    \label{fig_rtt_cdf}
\end{figure*}

On the other hand, the proposed online calibration method optimizes every parameter during the application operation.
Almost the same conditions from the previous experiments were used in this experiment.
For instance, we performed $B=8$ times ranging procedures for the initial calibration step and used a learning rate of 0.001 for both the initial and self-calibration procedures.
The initial values for the ranging parameters were given by $c_1=1$ and $c_0=0$.
In addition, the self-calibration procedure for the ranging module was executed every 30~s using up to 100 latest Wi-Fi ranging results.
Additionally, the same early stopping criteria were applied to both initial and self-calibration steps.

Fig.~\ref{fig_rtt_init} illustrates the initial calibration results for a selected device (i.e., Google Pixel 3).
As the initial calibration iteration increases, the parameters in the system are jointly optimized to minimize the cost, as shown in Fig.~\ref{fig_rtt_init}(a).
The initial calibration procedure was terminated with 1457 iterations, which took only 98~ms on the device.
The benchmark parameters and initial estimates of parameters using the proposed method are summarized in Table~\ref{table_rtt_parm}.

Fig.~\ref{fig_rtt_run} depicts an example of the online calibration procedure.
Fig.~\ref{fig_rtt_run}(a) illustrates the estimated trajectory of the device using the proposed method with the excessive Wi-Fi ranging scenario.
Note that the green area now indicates a 1~m error region.
Fig.~\ref{fig_rtt_run}(b) and (c) indicate that the ranging parameters are updated every 30~s, whereas the parameters in the PDR module are more frequently updated with the Wi-Fi ranging results.
Similar to the experiments for the RSS-based ranging scenario, each self-calibration procedure completed within a small number of iterations.
The average number of iterations and average computation time for the self-calibration procedure are also summarized in Table~\ref{table_rtt_parm}.

Fig.~\ref{fig_rtt_all} illustrates the estimated trajectories of every device for the benchmark scenario and proposed online calibration scenarios with $\rho=0$ and 0.8 scenarios.
Every trajectory in the benchmark scenario is precisely aligned to the test path in the initial stage.
Fig.~\ref{fig_rtt_all}(b) indicates that the proposed method with the excessive ranging scenario also produces accurate trajectories for all devices.
However, the estimated trajectories with $\rho=0.8$ produces more errors as the Wi-Fi ranging procedure is less frequently executed.

\begin{table}
\caption{Positioning Performance with RTT-based Ranging and PDR}
\label{table_rtt_result}
\renewcommand{\arraystretch}{1.2}
\centering
\begin{tabular}{c|c|cccc|c}
 & Device \# & 5 & 6 & 7 & 8 & Mean\\
\hline 
{\multirow{4}{*}{\rotatebox[origin=c]{90}{Benchmark}}} & Mean ranging interval [s] & 0.5  & 0.5  & 0.5  & 0.5  & 0.5\\
 & MAE [m] & 1.03 & 0.59 & 1.20 & 0.90 & 0.93\\
 & RMSE [m] & 1.18 & 0.67 & 1.32 & 0.97 & 1.04 \\
 & 75\%-tile error [m] & 1.49 & 0.83 & 1.69 & 1.12 & 1.28\\
 \hline
{\multirow{4}{*}{\rotatebox[origin=c]{90}{Proposed}}}{\multirow{4}{*}{\rotatebox[origin=c]{90}{$\rho=0$}}} & Mean ranging interval [s] & 0.5 & 0.5 & 0.5 & 0.5  &   0.5\\
 & MAE [m] & 1.27 & 1.13 & 0.81 & 0.94  &  1.04\\
 & RMSE [m] & 1.41 & 1.25 & 0.92 & 1.05  &  1.16\\
 & 75\%-tile error [m] & 1.67 & 1.58 & 1.01 & 1.32  &  1.39\\
 \hline
{\multirow{4}{*}{\rotatebox[origin=c]{90}{Proposed}}}{\multirow{4}{*}{\rotatebox[origin=c]{90}{$\rho=0.2$}}} & Mean ranging interval [s] & 1.01 & 1.89 & 1.95 & 1.64  &  1.62\\
 & MAE [m] & 1.39 & 1.70 & 1.91 & 1.10  &  1.52\\
 & RMSE [m] & 1.56 & 1.84 & 2.04 & 1.27  &  1.68\\
 & 75\%-tile error [m] & 1.83 & 1.92 & 2.65 & 1.72  &  2.03\\
 \hline
{\multirow{4}{*}{\rotatebox[origin=c]{90}{Proposed}}}{\multirow{4}{*}{\rotatebox[origin=c]{90}{$\rho=0.4$}}} & Mean ranging interval [s] & 3.65 & 7.90 & 8.47 & 6.96  &  6.74\\
 & MAE [m] & 1.42 & 1.80 & 2.09 & 1.37  &  1.67\\
 & RMSE [m] & 1.88 & 1.86 & 2.21 & 1.56  &  1.88\\
 & 75\%-tile error [m] & 2.17 & 2.19 & 2.64 & 2.21   &  2.30\\
 \hline
{\multirow{4}{*}{\rotatebox[origin=c]{90}{Proposed}}}{\multirow{4}{*}{\rotatebox[origin=c]{90}{$\rho=0.8$}}} & Mean ranging interval [s] & 16.6 & 23.7 & 25.4 & 24.7  &  22.6\\
 & MAE [m] & 2.64 & 1.57 & 1.78 & 1.59  &  1.89\\
 & RMSE [m] & 2.79 & 1.74 & 2.13 & 1.70  &  2.09\\
 & 75\%-tile error [m] & 3.14 & 1.93 & 2.84 & 2.09 & 2.50\\
 \hline
\end{tabular}
\end{table}

Fig.~\ref{fig_rtt_cdf}(a) depicts the MAE performance for every device with different ranging threshold values. 
Although a larger $\rho$ results in more accurate positioning performance for some scenarios, the general trend is that the MAE increases with the ranging threshold.
Fig.~\ref{fig_rtt_cdf}(b) indicates that the mean ranging interval increases with the ranging threshold.
Fig.~\ref{fig_rtt_cdf}(c) illustrates the CDF of positioning accuracy of the benchmark scenario and the proposed method with different $\rho$ values.
When the ranging procedure is executed at the same frequency as the benchmark scenario, the positioning accuracy of the proposed method closely approaches to that of benchmark performance, although the parameters are optimized in real-time.
In addition, the figure indicates that the CDF curve of the proposed method shifts to the right as a larger $\rho$ is applied.
Finally, the positioning performance with RTT-based ranging and PDR is summarized in Table~\ref{table_rtt_result}.

\section{Conclusion}

In this study, we investigated the online parameter calibration technique that optimizes every parameter in Wi-Fi ranging and PDR modules in real-time. 
The performance of the proposed method was extensively verified using various mobile devices for both RSS and RTT-based ranging scenarios.
Although the proposed method did not perform any prior calibration procedures, the positioning accuracy closely approached that of the benchmark performance, where every parameter was selected in an optimal manner.
We believe that the proposed method can be modified in various ways. For instance, the self-calibration of the ranging module can be skipped if the ranging module is correctly calibrated or, if the user visits a previously visited site, the ranging parameters can be initialized according to the last calibration results under the same site.
In addition, this paper also discussed an efficient way of performing irregular Wi-Fi ranging procedures to improve battery life and network performance. 
This method can also be modified more sophisticatedly, for instance, the application may initiate the Wi-Fi ranging procedure when the device encounters an unexpected situation.

\ifCLASSOPTIONcaptionsoff
  \newpage
\fi


\begin{thebibliography}{10}
\providecommand{\url}[1]{#1}
\csname url@samestyle\endcsname
\providecommand{\newblock}{\relax}
\providecommand{\bibinfo}[2]{#2}
\providecommand{\BIBentrySTDinterwordspacing}{\spaceskip=0pt\relax}
\providecommand{\BIBentryALTinterwordstretchfactor}{4}
\providecommand{\BIBentryALTinterwordspacing}{\spaceskip=\fontdimen2\font plus
\BIBentryALTinterwordstretchfactor\fontdimen3\font minus
  \fontdimen4\font\relax}
\providecommand{\BIBforeignlanguage}[2]{{%
\expandafter\ifx\csname l@#1\endcsname\relax
\typeout{** WARNING: IEEEtran.bst: No hyphenation pattern has been}%
\typeout{** loaded for the language `#1'. Using the pattern for}%
\typeout{** the default language instead.}%
\else
\language=\csname l@#1\endcsname
\fi
#2}}
\providecommand{\BIBdecl}{\relax}
\BIBdecl

\bibitem{4796924}
Y.~Gu, A.~Lo, and I.~Niemegeers, ``A survey of indoor positioning systems for
  wireless personal networks,'' \emph{IEEE Commun. Surveys Tuts.},
  vol.~11, no.~1, pp. 13--32, First Quarter 2009.

\bibitem{7039067}
L.~Mainetti, L.~Patrono, and I.~Sergi, ``A survey on indoor positioning
  systems,'' in \emph{Proc. 22nd Int. Conf. Software,
  Telecommun. Comput. Netw. (SoftCOM)}, Sep. 2014, pp. 111--120.

\bibitem{correa_sen_17}
A.~Correa, M.~Barcelo, A.~Morell, and J.~L. Vicario, ``A review of pedestrian
  indoor positioning systems for mass market applications,'' \emph{Sensors},
  vol.~17, no.~8, p.~1927, Aug. 2017.

\bibitem{8409950}
C.~{Laoudias}, A.~{Moreira}, S.~{Kim}, S.~{Lee}, L.~{Wirola}, and
  C.~{Fischione}, ``A survey of enabling technologies for network localization,
  tracking, and navigation,'' \emph{IEEE Commun. Surveys Tuts.},
  vol.~20, no.~4, pp. 3607--3644, Fourth Quarter 2018.

\bibitem{8852722}
V.~{Renaudin} \emph{et~al.}, ``Evaluating indoor positioning systems in a
  shopping mall: The lessons learned from the IPIN 2018 competition,''
  \emph{IEEE Access}, vol.~7, pp. 148,594--148,628, 2019.

\bibitem{8007254}
J.~{Jun}, L.~{He}, Y.~{Gu}, W.~{Jiang}, G.~{Kushwaha}, V.~{A.}, L.~{Cheng},
  C.~{Liu}, and T.~{Zhu}, ``Low-overhead WiFi fingerprinting,'' \emph{IEEE Trans. Mobile Comput.}, vol.~17, no.~3, pp. 590--603, Mar. 2018.

\bibitem{Yu_2019}
Y.~Yu, R.~Chen, L.~Chen, G.~Guo, F.~Ye, and Z.~Liu, ``A robust dead reckoning
  algorithm based on Wi-Fi FTM and multiple sensors,'' \emph{Remote Sensing},
  vol.~11, no.~5, p.~504, Mar. 2019.

\bibitem{8839041}
J.~{Choi}, Y.-S.~{Choi}, and S.~{Talwar}, ``Unsupervised learning techniques for
  trilateration: From theory to android APP implementation,'' \emph{IEEE
  Access}, vol.~7, pp. 134,525--134,538, 2019.

\bibitem{8911751}
K.~{Han}, S.~M. {Yu}, and S.~{Kim}, ``Smartphone-based indoor localization
  using Wi-Fi fine timing measurement,'' in \emph{Proc. Int. Conf.
  Indoor Positioning Indoor Navigat. (IPIN)}, Sep. 2019, pp. 1--5.

\bibitem{8924707}
G.~{Guo}, R.~{Chen}, F.~{Ye}, X.~{Peng}, Z.~{Liu}, and Y.~{Pan}, ``Indoor
  smartphone localization: A hybrid WiFi RTT-RSS ranging approach,'' \emph{IEEE
  Access}, vol.~7, pp. 176,767--176,781, 2019.

\bibitem{8854290}
Y.~{Zhao}, W.~{Wong}, T.~{Feng}, and H.~K.~{Garg}, ''Efficient and scalable calibration-free indoor positioning using crowdsourced data,'' \emph{IEEE Internet Things J.}, vol.~7, no.~1, pp.~160--175, Jan. 2020.

\bibitem{9043567}
X.~{Wang}, D.~{Qin}, R.~{Guo}, M.~{Zhao}, L.~{Ma}, and T.~M.~{Berhane}, ``The technology of crowd-sourcing landmarks-assisted smartphone in indoor localization,'' \emph{IEEE Access}, vol.~8, pp.~57,036--57,048, 2020.

\bibitem{Tian2014}
Z.~Tian, Y.~Zhang, M.~Zhou, and Y.~Liu, ``Pedestrian dead reckoning for MARG
  navigation using a smartphone,'' \emph{EURASIP J. Adv. Signal
  Process.}, vol. 2014, no.~1, p.~65, May 2014.

\bibitem{Diaz2015}
E.~M. Diaz, ``Inertial pocket navigation system: Unaided 3D positioning,''
  \emph{Sensors}, vol.~15, no.~4, pp. 9156--9178, Apr. 2015.

\bibitem{6987239}
W.~{Kang} and Y.~{Han}, ``SmartPDR: Smartphone-based pedestrian dead reckoning
  for indoor localization,'' \emph{IEEE Sensors J.}, vol.~15, no.~5, pp.
  2906--2916, May 2015.

\bibitem{6971168}
B.~{Zhou}, Q.~{Li}, Q.~{Mao}, W.~{Tu}, and X.~{Zhang}, ``Activity
  sequence-based indoor pedestrian localization using smartphones,'' \emph{IEEE
  Trans. on Human-Mach. Syst.}, vol.~45, no.~5, pp. 562--574, Oct. 2015.

\bibitem{Wang2018}
B.~Wang, X.~Liu, B.~Yu, R.~Jia, and X.~Gan, ``Pedestrian dead reckoning based
  on motion mode recognition using a smartphone,'' \emph{Sensors}, vol.~18, no.~6, p.~1811, Jun. 2018.

\bibitem{8455482}
A.~{Solin}, S.~{Cortes}, E.~{Rahtu}, and J.~{Kannala}, ``Inertial odometry on
  handheld smartphones,'' in \emph{Proc. 21st Int. Conf. Inf. Fusion (FUSION)}, Jul. 2018, pp. 1--5.

\bibitem{8385119}
H.~{Ju}, S.~Y. {Park}, and C.~G. {Park}, ``A smartphone-based pedestrian dead
  reckoning system with multiple virtual tracking for indoor navigation,''
  \emph{IEEE Sensors J.}, vol.~18, no.~16, pp. 6756--6764, Aug. 2018.

\bibitem{Zhang2018}
Y.~Zhang, F.~Yu, W.~Gao, and Y.~Wang, ``An improved strapdown inertial
  navigation system initial alignment algorithm for unmanned vehicles,''
  \emph{Sensors}, vol.~18, no.~10, p. 3297, Sep. 2018.

\bibitem{8756098}
H.~{Zhao}, L.~{Zhang}, S.~{Qiu}, Z.~{Wang}, N.~{Yang}, and J.~{Xu},
  ``Pedestrian dead reckoning using pocket-worn smartphone,'' \emph{IEEE
  Access}, vol.~7, pp. 91,063--91,073, 2019.

\bibitem{Xu2019}
L.~Xu, Z.~Xiong, J.~Liu, Z.~Wang, and Y.~Ding, ``A novel pedestrian dead
  reckoning algorithm for multi-mode recognition based on smartphones,''
  \emph{Remote Sensing}, vol.~11, no.~3, p.~294, Feb. 2019.

\bibitem{Liu2012}
J.~Liu, R.~Chen, L.~Pei, R.~Guinness, and H.~Kuusniemi, ``A hybrid smartphone
  indoor positioning solution for mobile LBS,'' \emph{Sensors}, vol.~12,
  no.~12, pp. 17,208--17,233, Dec. 2012.

\bibitem{6714451}
Z.~Yang, L.~Shangguan, W.~Gu, Z.~Zhou, C.~Wu and Y.~Liu, "Sherlock: Micro-environment sensing for smartphones," \emph{IEEE Trans. Parallel Distrib. Syst.}, vol.~25, no.~12, pp.~3295--3305, Dec. 2014.

\bibitem{6817916}
V.~{Radu} and M.~K. {Marina}, ``HiMLoc: Indoor smartphone localization via
  activity aware pedestrian dead reckoning with selective crowdsourced WiFi
  fingerprinting,'' in \emph{Proc. Int. Conf. Indoor Positioning Indoor Navigat. (IPIN)}, Oct. 2013, pp. 1--10.

\bibitem{7935650}
H.~{Zou}, Z.~{Chen}, H.~{Jiang}, L.~{Xie}, and C.~{Spanos}, ``Accurate indoor
  localization and tracking using mobile phone inertial sensors, WiFi and
  iBeacon,'' in \emph{Proc. IEEE Int. Symp. Inertial Sensors Syst. (INERTIAL)}, Mar. 2017, pp. 1--4.

\bibitem{8756276}
S.~{Xu}, R.~{Chen}, Y.~{Yu}, G.~{Guo}, and L.~{Huang}, ``Locating smartphones
  indoors using built-in sensors and Wi-Fi ranging with an enhanced particle
  filter,'' \emph{IEEE Access}, vol.~7, pp. 95,140--95,153, 2019.


\bibitem{Wang2003AnIW}
Y.-C. Wang, X.~Jia, and H.~K. Lee, ``An indoor wireless positioning system
  based on wireless local area network infrastructure,'' in \emph{Proc. 6th
  Int. Symp. Satell. Navigat. Technol. Including Mobile Positioning and Location Services}, Jul. 2003, pp. 1--13.

\bibitem{5425237}
J.~{Yang} and Y.~{Chen}, ``Indoor localization using improved RSS-based
  lateration methods,'' in \emph{Proc. IEEE Global Telecommun. Conf. (GLOBECOM)}, Nov. 2009, pp. 1--6.

\bibitem{5558044}
C.~{Shih} and P.~J. {Marrón}, ``COLA: Complexity-reduced trilateration
  approach for 3D localization in wireless sensor networks,'' in \emph{Proc. 4th Int. Conf. Sensor Technol. Appl.},
  Jul. 2010, pp. 24--32.

\bibitem{5766644}
B.~{Kim}, W.~{Bong}, and Y.~C. {Kim}, ``Indoor localization for Wi-Fi devices
  by cross-monitoring AP and weighted triangulation,'' in \emph{Proc. IEEE
  Consum. Commun. Netw. Conf. (CCNC)}, Jan. 2011, pp. 933--936.

\bibitem{832252}
P.~{Bahl} and V.~N. {Padmanabhan}, ``RADAR: an in-building RF-based user
  location and tracking system,'' in \emph{Proc. IEEE Int. Conf. Comput. Commun. (INFOCOM)}, vol.~2, Mar. 2000, pp. 775--784.

\bibitem{1047316}
P.~{Prasithsangaree}, P.~{Krishnamurthy}, and P.~{Chrysanthis}, ``On indoor
  position location with wireless LANs,'' in \emph{Proc. 13th IEEE Int. Symp. Personal, Indoor, Mobile Radio Commun. (PIMRC)}, vol.~2, Sep.
  2002, pp. 720--724.
  
\bibitem{horus05}
M.~Youssef and A.~Agrawala, ``The Horus WLAN location determination system,''
  in \emph{Proc. 3rd Int. Conf. Mobile Syst., Appl., Services (MobiSys)}, Jun. 2005, pp. 205--218.
 

\bibitem{7786995}
\emph{IEEE Standard for Information Technology--Telecommunications and Information
  Exchange Between Systems Local and Metropolitan Area Networks--Specific
  Requirements - Part 11: Wireless LAN medium access control (MAC) and physical
  layer (PHY) specifications}, IEEE Standard 802.11-2016 (Revision of IEEE
  Standard 802.11-2012), Dec. 2016, pp. 1--3534.

\bibitem{ibrahim_mobicom_18}
M.~Ibrahim, H.~Liu, M.~Jawahar, V.~Nguyen, M.~Gruteser, R.~Howard, B.~Yu, and
  F.~Bai, ``Verification: Accuracy evaluation of WiFi fine time measurements on
  an open platform,'' in \emph{Proc. 24th Annu. Int. Conf. Mobile Comput. Netw. (MobiCom)}, Nov. 2018, pp. 417--427.
  
\bibitem{8911824}
J.~{Choi}, Y.-S.~{Choi}, and S.~{Talwar}, ``Unsupervised learning technique to
  obtain the coordinates of Wi-Fi access points,'' in \emph{Proc. Int. Conf. Indoor Positioning Indoor Navigat. (IPIN)}, Sep. 2019,
  pp. 1--6.

\bibitem{Khairi_14}
K.~Abdulrahim, T.~Moore, C.~Hide, and C.~Hill, ``Understanding the performance
  of zero velocity updates in MEMS-based pedestrian navigation,''
  \emph{Int. J. Adv. Technol.}, vol.~5, no.~2, pp. 53--60, Mar. 2014.

\bibitem{4027597}
X.~{Li}, ``RSS-based location estimation with unknown pathloss model,''
  \emph{IEEE Trans. on Wireless Commun.}, vol.~5, no.~12, pp.
  3626--3633, Dec. 2006.

\bibitem{res_sen_11}
A.~Bel, J.~Lopez~Vicario, and G.~Seco-Granados, ``Localization algorithm with
  on-line path loss estimation and node selection,'' \emph{Sensors}, vol.~11, no.~7, pp. 6905--6925, Jul. 2011.

\bibitem{6509479}
M.~R. {Gholami}, R.~M. {Vaghefi}, and E.~G. {Ström}, ``RSS-based sensor
  localization in the presence of unknown channel parameters,'' \emph{IEEE
  Trans. Signal Process.}, vol.~61, no.~15, pp. 3752--3759, Aug. 2013.

\bibitem{ipin16}
L.~Banin, U.~Schatzberg, and Y.~Amizur, ``WiFi FTM and map information fusion
  for accurate positioning,'' in \emph{Proc. Int. Conf. Indoor
  Positioning Indoor Navigat. (IPIN)}, Oct. 2016. pp. 1--4.

\bibitem{intel17}
L.~Banin, O.~Bar-Shalom, N.~Dvorecki, and Y.~Amizur, ``High-accuracy indoor
  geolocation using collaborative time of arrival (CToA),'' \emph{Intel White
  Paper}, Sep. 2017.

\bibitem{8579543}
L.~{Banin}, O.~{Bar-Shalom}, N.~{Dvorecki}, and Y.~{Amizur}, ``Scalable Wi-Fi
  client self-positioning using cooperative FTM-sensors,'' \emph{IEEE
  Trans. Instrum. Meas.}, vol.~68, no.~10, pp. 3686--3698, Oct. 2019.

\bibitem{Zij_2004}
W.~Zijlstra, ``Assessment of spatio-temporal parameters during unconstrained
  walking,'' \emph{European J. Appl. Physiol.}, vol.~92, pp. 39--44, 2004.

\bibitem{5646888}
J.~{Jahn}, U.~{Batzer}, J.~{Seitz}, L.~{Patino-Studencka}, and J.~{Gutiérrez
  Boronat}, ``Comparison and evaluation of acceleration based step length
  estimators for handheld devices,'' in \emph{Proc. Int. Conf. Indoor Positioning Indoor Navigat. (IPIN)}, Sep. 2010, pp. 1--6.

\bibitem{Valenti2015}
R.~G. Valenti, I.~Dryanovski, and J.~Xiao, ``Keeping a good attitude: A
  quaternion-based orientation filter for IMUs and MARGs,'' \emph{Sensors},
  vol.~15, no.~8, pp. 19,302--19,330, Aug. 2015.

\bibitem{Feng2017}
K.~Feng, J.~Li, X.~Zhang, C.~Shen, Y.~Bi, T.~Zheng, and J.~Liu, ``A new
  quaternion-based Kalman filter for real-time attitude estimation using the
  two-step geometrically-intuitive correction algorithm,'' \emph{Sensors},
  vol.~17, no.~9, p. 2146, Sep. 2017.

\bibitem{Bernal-Polo2019}
P.~Bernal-Polo and H.~Mart{\'i}nez-Barber{\'a}, ``Kalman filtering for attitude
  estimation with quaternions and concepts from manifold theory,''
  \emph{Sensors}, vol.~19, no.~1, p. 149, Jan. 2019.
  
\bibitem{Diebel06}
J.~Diebel, ``Representing attitude: Euler angles, unit quaternions, and
  rotation vectors,'' 2006.

\bibitem{1275684}
K.~W. Cheung, H.~C. So, W.~. Ma, and Y.~T. Chan, ``Least squares algorithms for
  time-of-arrival-based mobile location,'' \emph{IEEE Trans. Signal
  Process.}, vol.~52, no.~4, pp. 1121--1130, Apr. 2004.

\bibitem{paula_sen_2011}
B.~A.~M. Tarrío, P. and J.~R. Casar, ``Weighted least squares techniques for
  improved received signal strength based localization,'' \emph{Sensors}, vol.~11, no.~9, pp. 8569--8592, Sep. 2011.

\bibitem{custom_scan}
\BIBentryALTinterwordspacing
 [Online]. Available: \url{https://github.com/wj2dy/CustomWiFiScan}
\BIBentrySTDinterwordspacing

\bibitem{demo_video}
\BIBentryALTinterwordspacing
 [Online]. Available: \url{https://youtu.be/_FnznD1gxV0}
\BIBentrySTDinterwordspacing

\end{thebibliography}
\end{document}